\documentclass[11pt]{article}
\usepackage{adjustbox}
\usepackage{amsmath}
\usepackage{amssymb}
\usepackage{dutchcal}
\usepackage{enumitem}
\usepackage{fancyhdr}
\usepackage{float}
\usepackage[T1]{fontenc}
\usepackage{graphicx}
\usepackage{hhline}
\usepackage[utf8]{inputenc}
\usepackage{mathbbol}
\usepackage{mathtools}
\usepackage{multicol}
\usepackage{multirow}
\usepackage[normalem]{ulem}
\usepackage{wasysym}
\usepackage[paperheight=29.7cm,paperwidth=21.0cm,left=2.35cm,right=2.35cm,top=2.5cm,bottom=2.0cm]{geometry}
\usepackage{xurl}
\usepackage{amsmath}
\usepackage{courier}

\newcommand\tab[1][1cm]{\hspace*{#1}}

\begin{document}

\begin{center}
{\Large \textbf{Pairing Conceptual Modeling with Machine Learning}}
\end{center}

\begin{center}
Wolfgang Maass\textsuperscript{1,2}, Veda C. Storey\textsuperscript{3}
\end{center}

\begin{center}
\textsuperscript{1}German Research Center for Artificial Intelligence (DFKI), 66123 Saarbrücken, Germany
\end{center}

\begin{center}
\textsuperscript{2}Saarland University, Saarland Informatics Campus, 66123 Saarbrücken, Germany, wolfgang.maass@dfki.de
\end{center}

\begin{center}
\textsuperscript{3}J. Mack Robinson College of Business, Georgia State University, \\  Atlanta, GA 30302-4015 U.S.A.\\
vstorey@gsu.edu
\end{center}

\begin{center}
{\large \textbf{Abstract}}
\end{center}

Both conceptual modeling and machine learning have long been recognized as important areas of research. With the increasing emphasis on digitizing and processing large amounts of data for business and other applications, it would be helpful to consider how these areas of research can complement each other. To understand how they can be paired, we provide an overview of machine learning foundations and development cycle. We then examine how conceptual modeling can be applied to machine learning and propose a framework for incorporating conceptual modeling into data science projects. The framework is illustrated by applying it to a healthcare application. For the inverse pairing, machine learning can impact conceptual modeling through text and rule mining, as well as knowledge graphs. The pairing of conceptual modeling and machine learning in this this way should help lay the foundations for future research.

\textbf{Keywords}: Conceptual modeling, machine learning, methodologies and tools, models, database management, Framework for incorporating conceptual modeling into data science projects, artificial intelligence

\section{Introduction}

Machine learning (ML) emerged several decades ago as part of research in Artificial Intelligence (AI) and has recently received a surge in interest due to the increased digitization of data and processes. Machine learning uses data and algorithms to build models that carry out certain tasks without being explicitly programmed \cite{g16, lecun2015deep}. While machine learning focuses on technologies, its application is part of data science. More precisely, data science uses principles, processes, and techniques for understanding phenomena via analysis of data \cite{pf13}. Although data science could actually be performed manually (pen and pencil, or by calculators), the real power of data science becomes apparent by leveraging machine learning on big data. Data science supports data-driven decision making \cite{bm16} and is a key driver for digital transformation \cite{v19}.

Managing data, whether big or traditional, cannot be accomplished solely by humans with their limited cognitive capabilities. Rather, machine learning is important to address many business and societal problems that involve the processing of data. Machine learning has impacted many research fields, including natural sciences, medicine, management, and economics, and even humanities. In contrast to traditional software system development, machine learning does not require programming based on a given design, but rather requires fitting of parameters of generic models on data until the output (predictions, estimates, or results) minimizes or maximizes an objective function \cite{g16}. Many types of models have been developed and continue to be applied. Machine learning requires an in-depth understanding of the domains to which machine learning models and algorithms can be applied because data determines the functionality of an information system. Therefore, an assessment is required of whether the training data is representative of the domain. Otherwise, problems might arise that could contribute to biases and mistakes in machine learning models, of which there are well-documented examples, such as automatic parole decisions \cite{r19} or accidents with self-driving cars \cite{s18}.

Although machine learning continues to be an important part of business and society, there are many challenges associated with the progression machine learning so that it is increasingly accessible and useful. At the same time, the development of information systems, of any kind, first requires understanding and representing the real world, which, traditionally, has been the role of conceptual modeling. Emphasis on both big data generation and traditional applications highlights the need to understand, model, and manage data. Over the past decade, research has included the role of conceptual modeling on big data, business, healthcare, and many other applications  \cite{el13}. Conceptual modeling adds a perspective that starts with strategic business goals and finds translations and abstractions that finally guide software development \cite{pm07,a10, mv12}.

The purpose of this paper is to examine how conceptual modeling and machine learning can, and should, be combined to mutually support each other and, in doing so, improve the use of and access to machine learning. We make several contributions. First, the paper provides an overview of the foundations and development cycle of machine learning. Second, we derive a framework for incorporating conceptual modeling into data science projects and demonstrate its use through an application to a specific healthcare project. Third, we examine how machine learning can contribute to conceptual modeling activities. Suggested areas of research are also proposed. This paper shall help researchers and practitioners of conceptual modeling integrate machine learning into their research and operations while also helping data scientists and machine learning experts to use conceptual modeling in their work.

The paper proceeds as follows. Section 2 provides a brief summary of conceptual modeling and its potential use in machine learning. Section 3 presents the foundations of machine learning and its development cycle. Section 4 proposes a framework for incorporating machine learning into data science projects, which it applied to a health care problem. Section 5 highlights the potential impacts of machine learning for conceptual modeling. Section 6 outlines additional research directions for pairing these two fields. Section 7 concludes the paper.

\section{Conceptual Modeling and Machine Learning Pairing}

Conceptual modeling is described as $``$the activity for \textit{formally} describing some aspect of the physical and social world around us \textit{for the purposes of understanding and communication}$"$ (p. 51) \cite{m92}. Conceptual models attempt to capture requirements with the purpose of creating a shared understanding among various people during the design of a project within the boundaries of the application domain or an organization \cite{msk11}. They help to structure reality by abstracting the relevant aspects of a domain, while ignoring those that are not relevant. A conceptual model formally represents requirements and goals. It is shaped by the perspectives of the cognitive agents whose mental representations it captures. In this way, a conceptual model can  serve as a \textit{social artifact} with respect to the need to capture a shared conceptualization of a group \cite{ggm20}. Much research that attempts to understand and characterize research on the development and application of conceptual modeling (e.g., \cite{mt20,l20, p16, d18, lps18, c21}).

The field of conceptual modeling has evolved over the past four decades and has been influenced by many disciplines including programming languages, software engineering, requirements engineering, database systems, ontologies, and philosophy. Conceptual modeling activities have been broadly applied in the development of information systems over a wide range of domains for varied purposes \cite{d18}. Activities and topics related to conceptual modeling have evolved over the past four decades \cite{l20, hf20}.  Notably, Jaakkola and Thalheim \cite{jt21} highlight the importance of modeling, especially with the current emphasis on the development of artificial intelligence (AI) and machine learning (ML) tasks. Other research has also proposed the need for conceptual modeling to support machine learning and, in general, combining conceptual modeling with artificial intelligence \cite{ggm20, mt20, l20, p16, d18}.

Conceptual models are a $``$lens$"$ through which humans gain an $``$intuitive, easy to understand, meaningful, direct and natural mental representation of a domain$"$ \cite{ggm20}. In contrast, machine learning uses data as a $``$lens$"$ through which it gains internal representations on the regularities of data taken from a domain (\cite{htf09, jm15, d12}). Pairing conceptual modeling with machine learning contributes to each other by: 1) improving the quality of ML models by using conceptual models during data engineering, model training and model testing; 2) enhancing the interpretability of machine learning models by using conceptual models; and 3) enriching conceptual models by applying ML technologies.

\begin{figure}[H]
\includegraphics[width=16.29cm,height=6.65cm]{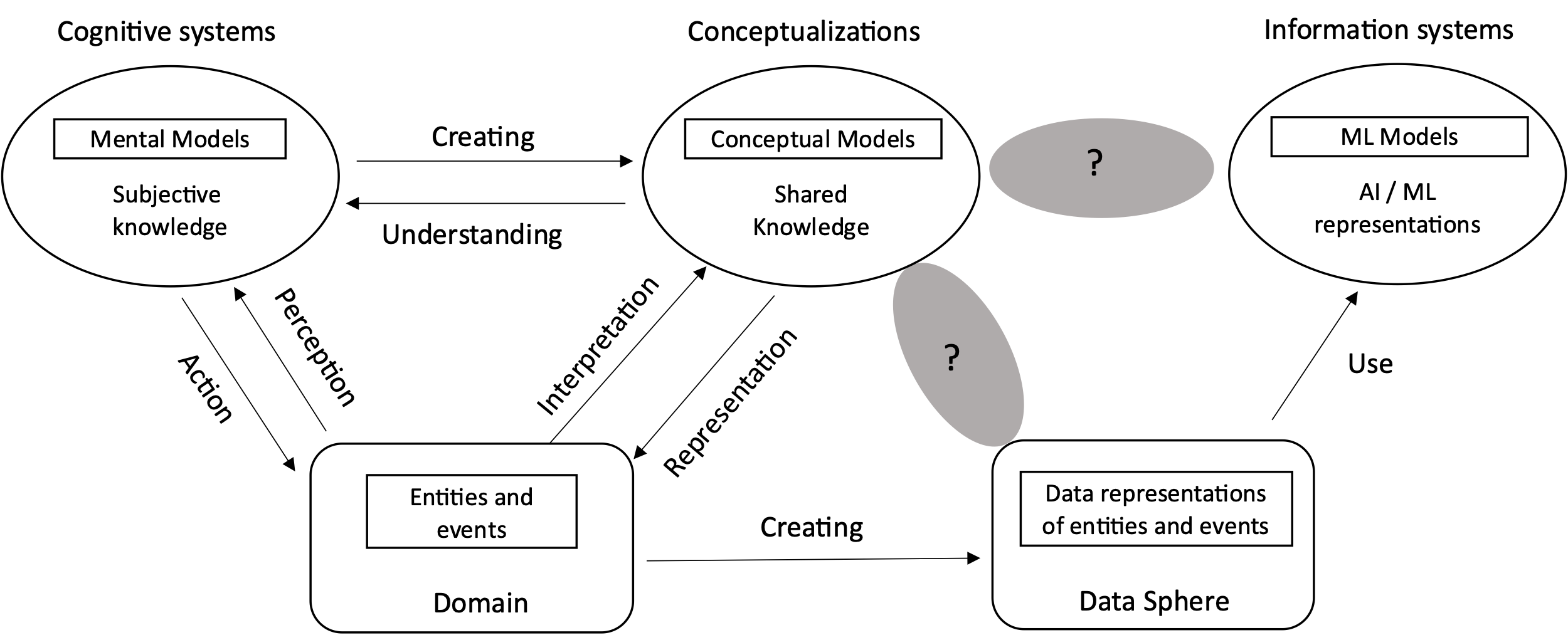}
\caption{Progression from mental models to Artificial Intelligence / Machine Learning models .}
\label{fig:framework}
\end{figure}

Figure \ref{fig:framework} summarizes the relationships among mental models, conceptual models, and machine learning (ML) models. Mental models naturally evolve by acting in domains, whereas conceptual models are shared conceptualizations of mental models \cite{n83}. For information system development, conceptual models represent shared conceptualization about a domain by means of conceptual modeling grammars and methods in given contexts \cite{ww02}. For information systems based on programming approaches, conceptual models are used as requirements for implementations. Database systems are designed and realized according to requirements expressed by conceptual models, such as entity-relationship models \cite{c76}. For learning-based information systems, relationships between data and conceptual models and ML models and conceptual models are less obvious \cite{msl21} (Figure \ref{fig:framework}).

\vspace{2\baselineskip}


In this paper, we examine these relationships in both directions.

\begin{enumerate}[label*=\arabic*.]
	\item How can conceptual modeling support the design and development of machine learning solutions?

	\item How can machine learning support the development and evaluation of conceptual models?

\end{enumerate}
Machine learning systems, which are applied to individual datasets, grow exponentially in both size and complexity. Conceptual modeling is instrumental in dealing with complex software development projects. Therefore, the first question to consider is whether conceptual modeling can help structure machine learning projects, create a common understanding, and thereby increase the quality of the resulting machine learning-based system. The second question reverses the direction and asks whether machine learning can provide tools that could support the development of conceptual models.  Then, it would also be important to assess how accurately a conceptual model captures an application domain. This is especially challenging for automatic validation, but could take advantage of a data-driven approach to augmenting conceptual modeling.

For information systems that depend on very large datasets and increasingly complexity, machine learning systems, biases of data and uncertainty in decision-making pose threats to trust, especially when the systems are used for recommendations. Pairing machine learning and conceptual modeling thus becomes an attempt to support \textit{Fair Artificial Intelligence} \cite{bs16}. This requires using structures and concepts during AI-based software development lifecycles that are stable and meaningful, yet have well-defined semantics, and can be interpreted by humans. We, therefore, propose that the role of conceptual modeling with respect to machine learning is (CM ML):

\begin{itemize}
	\item \textit{Descriptive}: informs data scientists when developing machine learning systems

	\item \textit{Computational}: embedded into machine learning implementations.

\end{itemize}
Inversely, the role of machine learning with respect to conceptual modeling is (CMML):

\begin{itemize}
	\item \textit{Descriptive}: informs conceptual modelers

	\item \textit{Computational}: constraints or creates conceptual models.

\end{itemize}
There are several ways in which conceptual models should be able to inform a data scientist’s work. Conceptual models provide conceptual semantics for concepts and relationships of a domain that govern data used for machine learning tasks. This knowledge can inform data scientists, especially during data engineering but also during model training, and model optimization.   If conceptual models are expressed by some computational modeling language, they can be integrated into ML models development procedures. For example, conceptual models can be used to derive constraints on data features that are automatically evaluated during data engineering. That makes both the descriptive and computational perspectives important. Inversely, regularities found by ML models can provide insights for conceptual modelers that can be used for revision and refinement of conceptual semantics and conceptual models. Thus, conceptual models and ML models are independent means for understanding domains of interest. Conceptual models that are consistent with ML models, and vice versa, can increase trustworthiness. Inconsistencies can be indicators of flaws in conceptual models or ML models, but might also facilitate the extraction of novel insights.

\section{Machine Learning: An Overview}

This section provides the foundations necessary to understand machine learning.

\subsection{Model Foundations}

Machine learning is part of data science projects where the results obtained from software are integrated into an information system, which is called a \textit{ML-based information system}. A data scientist is required to understand statistics and communicate and explain the design of a machine learning system to stakeholders of a data science project \cite{dp12}. Work on conceptual modeling is similar to that of  a storyteller who wants to communicate about the digital and real worlds. With the increasing complexity of ML-based information systems, future data scientists should also understand: how conceptual models can govern data and ML-based information systems; how to carry out conceptual modeling activities; and how to apply knowledge representations techniques.

Machine learning involves creating a \textbf{model }that  is trained on a set of \textbf{training data }and is then applied to additional data to make \textbf{predictions}. Various \textbf{types of models} have been used and researched for machine learning based systems \cite{htf09}. For example, a predictive model is a function of the form$f:X^{\ast }\rightarrow Y^{\ast }$ where$X^{\ast }$ represents a multidimensional input set and$Y^{\ast }$ a multi-dimensional output set. Every parametric model is defined by a set of parameters (aka weights)$w\in W$and applied to values of an input vector$x$. For instance, a simple linear regression has the form:

\begin{equation}
f\left(w,x\right) = w^{T}x = w_{0} + \sum_{i\in 1,\ldots ,n}^{}w_{i}x_{i}
\end{equation}

Values of$w$are derived (or $``$learned$"$) from a set of combinations$(x^{i},y^{i})$ with$x^{i} = (x_{1}^{i},\ldots ,x_{n}^{i})$ the i-th input vector from the input dataset$X$ ($ x^{i}\in X$), where$y^{i}$ is i-th output from the output dataset$Y$($ y^{i}\in Y$). Each$x_{j}^{i}\in X$ represents an attribute of an entity of interest, and  is\ \ called a \textit{feature}, e.g., age of a person or pixel of an image. For tabular data, a feature is a column. The task of \textit{supervised learning} is to determine the weight vector$w =\left(w_{1},\ldots , w_{n}\right)^{T}$in such a way that the difference (loss)$L(f\left(w,x\right),y)$ between an estimation of output vector $\hat{y} = f(w,x)$ given a new input vector$x$ and the actual value of$y$ (often called the \textit{ground truth}\footnote{Ground truth $Y$ is subject to uncertainties about its actual truthfulness. It only states that, for the modeling task, it is assumed that each $ y\in Y$ is true, with a probability $\alpha$ that this assumption is false. Often $\alpha$ is unkown.}) is minimized. Here, a loss function $L$ is used to measure how accurate a model is with weights$w$. An often-used loss functions is square loss:

\begin{center}
$ L\left(f\left(w,x\right),y\right) =\frac{1}{2}\left\Vert f\left(w,x\right) - y\right\Vert_{2}^{2} =$ $\frac{1}{2}\sum_{i\in 0\ldots n}^{}(f\left(w_{i},x_{i}\right) - y_{i})^{2}$
\end{center}

A loss function is a central element for machine learning algorithms because it is used for model optimization; that is, loss minimization. A loss function is also called an error function, cost function, or objective function. The latter name emphasizes its use as a criterion for optimizing a model \cite{g16}. Some ML models are designed to make finding optima tractable, such as linear regression and support vector machines. For others, such as deep learning models, finding a global maximum or minimum is intractable \cite{hl13}. Various heuristics, such as momentum or randomization, are used in combination with gradient descent algorithms to avoid being too restricted to a local minimum. Features of$X$ with small weights relative to other weights contribute little to the outcome. Sometimes prediction accuracy is improved by setting some weights to zero (cf. \cite{htf09}). Some methods start with the simple model consisting of the bias weight$w_{0}$ and only add dimensions with the highest impact on the prediction until the loss stabilizes.

\vspace{1\baselineskip}
\textbf{Model shrinkage methods}

Several reasons exist for reducing the complexity of a ML model. One reason is that it is important to identify which input variables are most important and have the strongest impact on predictions. This can be achieved by shrinking weights for variables with small to zero impact.  Another reason is that the number of input variables is larger than the sample size which generally results in perfect model fit with no noise. In this case, shrinking a model contributes to model generalization.

A model that uses all parameters by setting$w_{i}$ to zero has the least impact on model accuracy until the loss is stabilizing (cf. \cite{htf09}). Model shrinkage methods simultaneously adjust all weights$w_{i}$ by optimizing a risk function$R$ that is defined by the loss function$L(f\left(x\right), y)$ and an additional function$J\left(f\left(x\right)\right)$ that defines a penalty on model complexity. Therefore, a risk function$R(f)$ defines a tradeoff between minimizing loss and model complexity. The aim of a learning algorithm is to find a function$f:X\rightarrow Y $among a class of functions$F $for which the risk$R(f)$ is minimized:

\begin{equation}
f^{\ast } = argmin_{f\in F}R(f)
\end{equation}

where$f^{\ast }$ gives the best expected performance for loss$L(f\left(x,w\right),y)$ over the distribution of any function in$F$. Since the data generating distribution$P(X,Y)$ is unknown, the risk$R(f(w,x))$ cannot be computed. Therefore, minimizing the risk over a training dataset$X$\textbf{ }drawn from$P\left(X,Y\right)$ is formalized as:

\begin{equation}
R_{emp}\left(f(x,w)\right) =\frac{1}{n}\sum_{i\in 1\ldots n}^{}L\left(f\left(x_{i},w\right),y_{i}\right) + \lambda J(f(x_{i},w))
\end{equation}

With a standard definition of$J(f(x,w))$ as a norm:

\begin{center}
$ J\left(f(x,w)\right) = \sum_{i\in 1\ldots m}^{}\vert w_{i}\vert^{p}$ and$\lambda$ regularization parameter.
\end{center}

L1 ($ p = 1$) and L2 ($ p = 2$) norms are often used.$J\left(f(x,w)\right)$ is called a \textit{regularization function} because it controls model complexity. Using an L1 norm is called a \textit{lasso regression}, whereas using a L2 norm is called a \textit{ridge regression}. Several alternative regularization functions exist, such as\textit{ elastic net }\cite{zh05} and \textit{least angle regression }\cite{e04}. In contrast to reducing model complexity by principal components regression, model parameters restrict direct correspondence to input features. Thus, models keep their ability to directly support model explanations.

\subsubsection{Prediction types}

{\large Classification and regression supervised learning tasks are similar in that they both have numerical input but differ in the type of target variable being predicted. The numerical form is transformed based on a dependent variable that estimates output $y$ based on input vector $x$. Different types of target variable that determine different learning tasks are explained in the following sections.\par}

\vspace{1\baselineskip}
\textbf{Classification}

For classification, $f(x)$ projects input $x$ on categorical values given by a discrete dimension of $y$. The following types of classifications are used:

\begin{enumerate}[label*=\arabic*.]
	\item \textit{Binary Classification}: For every input $x$, there are only two possible output values for $y$.

	\item \textit{Multi-class Classification}: For every input $x$, there are more than two possible output values $y$.

	\item \textit{Multi-label Classification}: For every input $x$, there are more than two target values with each input sample associated with one or more labels, i.e.$y\in Y^{K}$with$K$ being the set of all class labels.

\end{enumerate}
An example for a binary classification is a binary logistic regression used for finding a function $f\left(x\right)$ that separates positive from negative illness cases.

\begin{equation}
p\left(x\right) =\frac{1}{1 + e^{ - (\beta_{0} + \beta_{1}X_{1})}}
\end{equation}

The logistic function provides the probability that the estimate of class $k$ is correct given input $x$ with $k\in \{ 0,1\}$. Figure \ref{fig:logreg} shows that $p\left(x\right)$ perfectly separates people younger than 25 years, and older than 26, with several misclassifications made in-between. The logistic function separates two topological spaces with associated labels $k\in \{ 0,1\}$ and selects the class with the highest probability.

\begin{center}
{\large  }\begin{figure}[H]
\centering
\includegraphics[width=10.22cm,height=6.34cm]{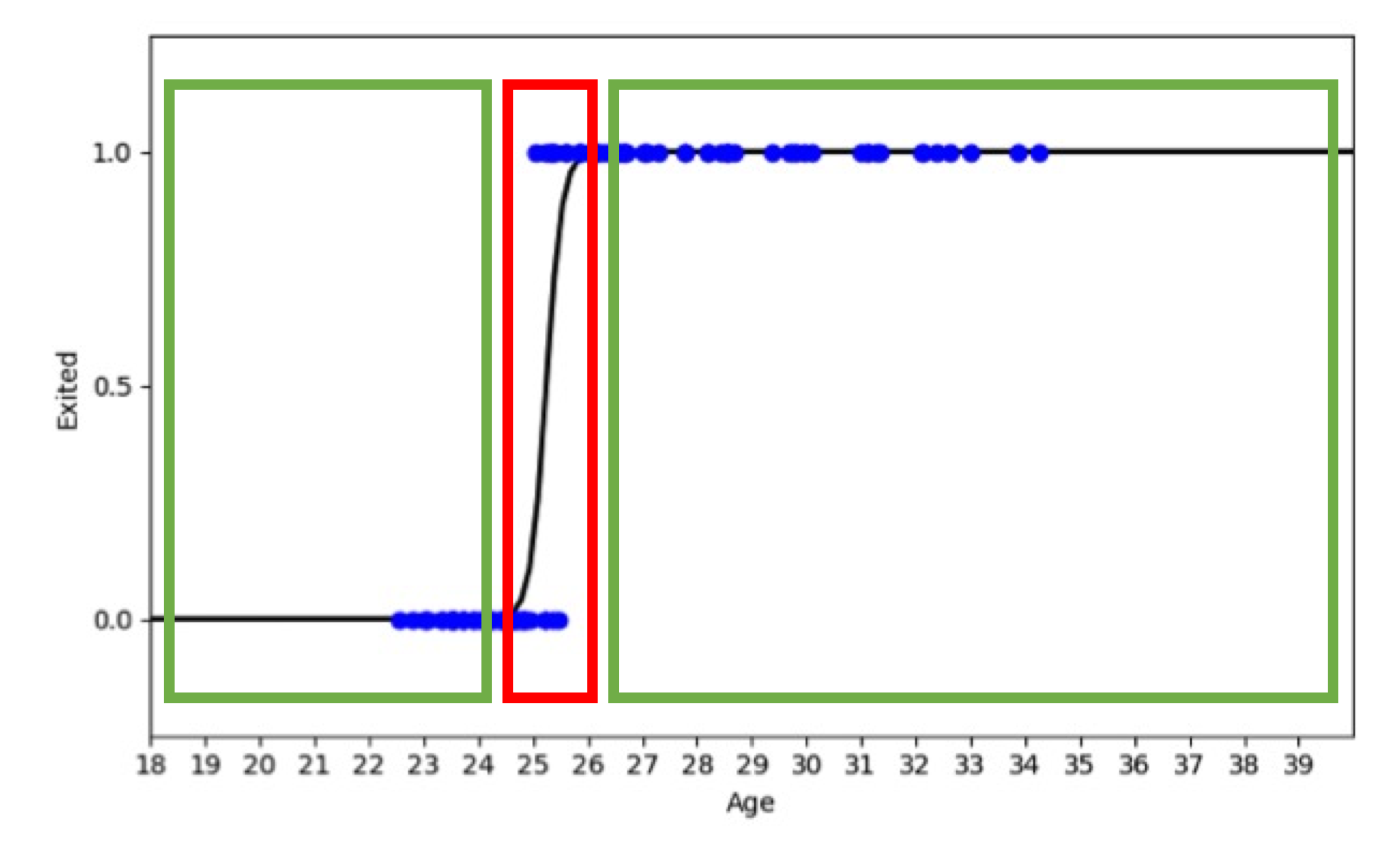}
\caption{Classification by using a binary logistic regression function; orange line: separation line of two classes.}
\label{fig:logreg}
\end{figure}

\end{center}


%
\vspace{1\baselineskip}
\textbf{Regression}

For regression, $f(x)$ projects input $x$ on numerical values given by the numerical dimension of $y$. Often the dimension of $y$ is real values ($ y\mathbb{\in R}$). Examples are estimates for stock prices or number of people swimming in a pool per day.

\subsubsection{Model selection}

Machine learning has a long history, dating as far back as the beginning of AI when the 1956 Dartmouth Summer Research Project contained \textit{neuron nets} as well as other topics. Because machine learning is built on mathematical statistics, the terms \textit{statistical learning} and \textit{machine learning} are often used synonymously \cite{jm15}. Given this long history, it is not surprising that a large variety of machine learning model types are available, which can be generally classified as follows.

\begin{itemize}
	\item \textit{Supervised learning}:  learning a function that maps an input to an output based on example input-output pairs \cite{rn10}.

	\item \textit{Unsupervised learning}: learning a function that maps an input to an output without prior labels available for output variables. Example models are k-nearest neighbors (KNN), k-means and clustering models.

	\item \textit{Reinforcement learning}: learning a function on how to take actions in an environment in order to maximize the notion of cumulative reward \cite{sb18}. Examples are Q-learning \cite{sb18}, Monte-Carlo \cite{sb18} or DQN \cite{m15}.

\end{itemize}
Recent approaches for unsupervised learning make use of reinforcement learning models so that the difference between both classes becomes less strict \cite{s17}. In the following, we focus on supervised learning, the most common form.

\vspace{1\baselineskip}
\textbf{\textbf{Supervised Learning }}

Commonly used model types for supervised learning are decision trees, ensemble learning and neural networks, each of which is discussed briefly below.

\vspace{1\baselineskip}
\textbf{\textit{Decision Trees}}

\begin{figure}[H]
\includegraphics[width=16.29cm,height=5.94cm]{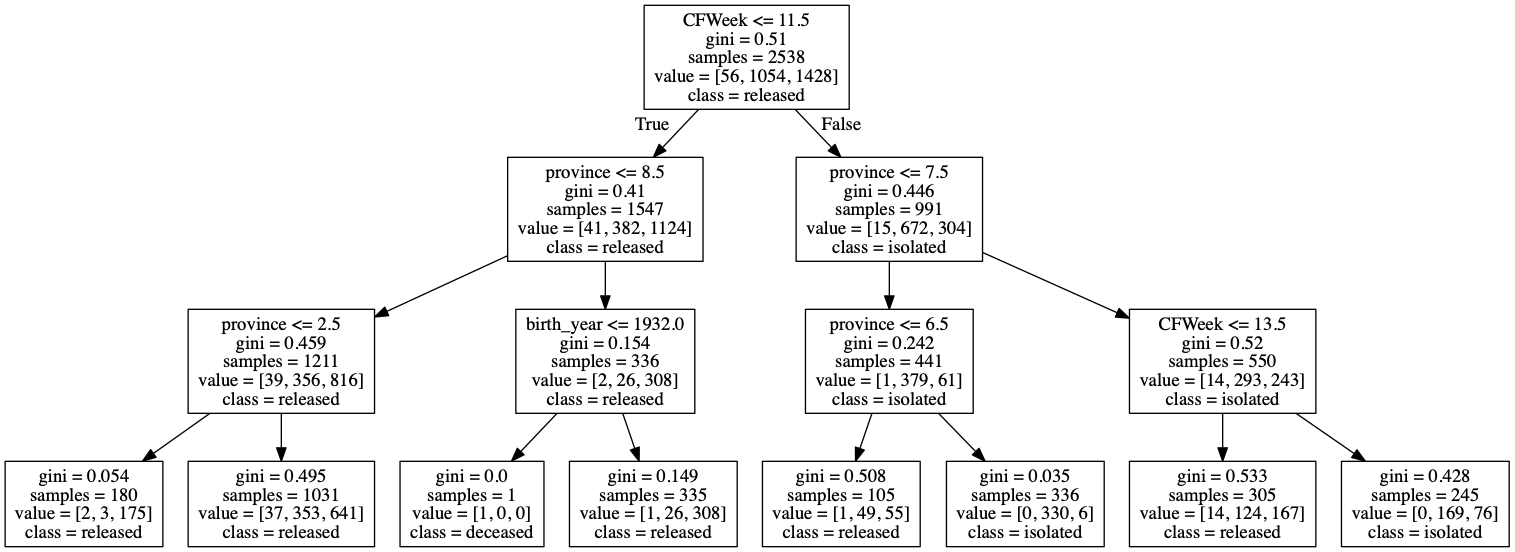}
\caption{Korean Covid-19 disease data.}
\label{fig:korea}
\end{figure}

Decision trees are widely used machine learning models. Particularly important are additive combinations of so-called \textit{weak learners}, such as boosting (AdaBoost \cite{f01} and XGBoost \cite{cg16}).  An early decision tree model is the Classification and Regression Trees (CART) model \cite{b84}. In the example (cf. Figure \ref{fig:korea}), a patient who was infected before week 11.5, in a province labeled smaller than 2.5 (i.e. provinces Busan, Chungcheongbuk-do and Chungcheongnam-do) is classified as being released (data source: \cite{kim}).

Decision trees are based on recursive decomposition mechanisms that optimize on a separation step. For classification, in each node with a data set$N_{i}$, the dimension$d\in D$ is selected that best separates data in cell$N_{i}$ according to a loss function$L(.)$. The most basic approach is to systematically consider one attribute after another, use any value of the dataset as threshold, and assess the loss$L(.)$. Loss functions are used for measuring the \textit{impurity} of resulting nodes after a split. For categorization, a node has a low impurity if it contains many samples from one category and few from others.

Typical loss functions \cite{htf09} are defined by using class probabilities $\hat{p}_{mk} = 1/N_{m}\sum_{x_{i}\in R_{m}}^{}I(y_{i} = k)$, with a node$m$ holding a region$R_{m}$ with$N_{m}$ data samples \cite{htf09}:

\begin{itemize}
	\item \textit{Gini index}:$\sum_{k\neq k'}^{K}\hat{p}_{mk}\hat{p}_{mk'} = \sum_{i = 1}^{K}\hat{p}_{mk}\left(1 -\hat{p}_{mk}\right)$for all classes$k\in K$
	\item \textit{Cross-entropy}: $-\sum_{i=1}^{K}\hat{p}_{mk}log(\hat{p}_{mk})$ for all classes $k\in K$.

\end{itemize}
Node$N_{0}$ contains 2538 patients of which 56 are labeled $``$deceased$"$, 1054 as $``$isolated$"$ and $``$1428$"$ as $``$released$"$. After reformulation of the equation of the Gini index, impurity of node$N_{0}$ calculates as follows:$G\left(N_{0}\right) = 1 -\left(\frac{56}{2538}\right)^{2} -\left(\frac{1054}{2538}\right)^{2} -\left(\frac{1428}{2538}\right)^{2} = 0.51$.

\vspace{2\baselineskip}


%
The decision criterium for a split is the impurity of resulting nodes that is calculated as weighted Gini impurity$I_{G}$. For the example, $I_{G}\left(N_{0}\right) =\frac{1547}{2538}\ast 0.41 +\frac{992}{2538}\ast 0.446 = 0.424$. By calculation of $I_{G}$, a feature and a threshold are selected resulting in the smallest lost, ie. smallest $I_{G}$.

Figure \ref{fig:korea} shows a decision tree derived from data of COVID-19 infection cases in South Korea trained for three categories ("deceased", "isolated", "released") based on eleven input variables, including year of birth, sex, age and confirmed day and month of infection.

\vspace{1\baselineskip}
\textbf{\textit{Ensemble Learning}}

Ensemble learning integrates base learners, such as decision trees, into complex models \cite{sr18}. The advantage of ensemble learning is in improved accuracy for the additional cost of increased latency due to testing a series of models at runtime and lack of interpretability. Instead of independent decision trees, \textit{boosting} iteratively integrates decision trees by focusing on mis-classified samples. For instance, \textit{AdaBoost} is a popular boosting model that only uses decision trees with one node and two leaves, called \textit{stumps }\ \ (\cite{fs97}). In each step, all samples of a dataset are associated with an equal weight$w_{i} =\frac{1}{number of samples}$. The attribute with the smallest Gini index is selected for the next stump. For each stump, the percentage of errors is computed:$error_{m}$ is the sum of weights$w_{i}$ misclassified divided by all weights$w_{i}$.

Additionally, each stump $m$ has a weight $\alpha_{m}=log(\frac{1-error_{m}}{error_{m}})$ that represents the importance of a stump. Small $\alpha_{m}$ means that a stump has little impact on the final result. From a stump, weights for each sample are updated according to a correct classification result: misclassified samples: $w_{i} = w_{i}\ast e^{\alpha_{m}}$ and correct classified samples:  $w_{i} = w_{i}\ast e^{ - \alpha_{m}}$. Then, weights for misclassified samples are increased and decreased for correctly classified samples. A new dataset is determined by drawing from samples with replacement by using normalized weights as probabilities; that is, misclassified samples have a higher chance of getting drawn multiple times. This new dataset resets all weights to equal weights again and the process restarts until the maximum number of iterations $M$ is reached.

\textit{Gradient boosting} extends the idea of AdaBoost by allowing larger trees with fixed size than stumps \cite{f01}. Gradient boosting starts with the mean$\gamma$ of the dependent variable that minimizes the loss function$L\left(y_{i}, \gamma\right) =\frac{1}{2}(y_{i} -\overline{y})^{2}$. Next, the so-called \textit{pseudo residuals} of each sample are determined which are the differences of the sample value and mean $\overline{y}$. Formally, the pseudo residual is the derivative of the squared loss function$L$which is$ -\frac{\partial L\left(y_{i}, f\left(x_{i}\right)\right)}{\partial f\left(x_{i}\right)} = f(x_{i}) - y_{i}$ with$f\left(x_{i}\right) = \gamma$ \cite{htf09}. Instead of growing a tree$m$on the dependent variable, it is grown on predicting the pseudo residuals. Thus, previous sample errors are corrected to a certain extend. For each leaf of terminal node$j,$ a value$\gamma_{jm}$ is determined that is the minimum of the loss function over all samples in this terminal node. The final estimate is the summation of all$\gamma_{jk}$ of regions where the input is associated with all trees of the gradient boosting model. When computing the estimate, the contribution of each tree is scaled by a learning rate$\eta$ for each tree result that controls for overfitting. Gradient boosting also supports classification by converting class labels into probabilities by applying the logistic function on the log of relative occurrence of classes in the dataset. The pseudo residuals are the differences between data values; the mean probability minimizes the negative log likelihood function used as loss function \cite{f01}.

\textit{XGBoost} is an extension of gradient boosting and optimization on the residuals \cite{cg16}. It uses an alternative for calculating the gain of a split based on the squared sum of residuals. A subtree is pruned if the gain is smaller than a threshold value$\gamma$. Regularization is achieved by another hyperparameter$\lambda$ used for decreasing gain values and, thus, the size of trees. Hence, XGBoost has two hyperparameters for controlling model complexity. Estimations are calculated such as gradient boosting controlled by a learning rate$\varepsilon$. XGBoost is designed for parallelization that is useful for large datasets. For classification, gain is computed by sample probabilities, similar to gradient boosting.

\vspace{1\baselineskip}
\textbf{\textit{Neural Networks}}

Neural networks represent a general class of learning models that can be adapted to different problems. For instance, convolutional neural networks (CNN) for visual computing (e.g., ResNet \cite{h16} and transformer models for natural language processing (e.g., BERT \cite{devlin18}). Neural networks make use of parallel execution of weak learners and, thus, train universal approximators of any function given sufficient data and resources \cite{h89}. Various forms of neural nets, such as recurrent neural nets (RNN) are \textit{Turing complete }\cite{ss95}.\textsuperscript{ }The proof of Turing completeness of more sophisticated model types, such as Transformer and Neural GPU \cite{pmb19}, is built on foundational mechanisms; that is, residual connections for Transformer and gates for Neural GPUs. These results provide evidence for the hypothesis that all, non-trivial machine learning model types are Turing complete. Decision trees and ensemble models based on decision trees lack the concept of loops and memory, so they are not Turing complete. However, the class of all machine learning model types as a whole, is Turing complete. Therefore, defining a model architecture is not a question of the complexity classes of computable functions, but of performance. A single node, called a neuron, consists of an application of a non-parametric non-linear function$\sigma$, called an \textit{activation function}, on a linear function with weights$w_{j}$:

\begin{equation}
\hat{y} = \sigma (\sum_{j = 1}^{p}w_{j}x_{j} + b)
\end{equation}

A neural network model is trained by fitting weights so that a corresponding loss function is minimized. Similar to gradient boosting, optimization of the loss function means to minimize residuals. Optimization of loss with respect to weights of the neural network is typically performed using a form of a gradient descent, such as gradient descent with momentum, or more sophisticated variants, such as the Adam optimization algorithm \cite{kb14}.

For classification tasks, a softmax function transforms output of the final activation layer to valid probabilities for each class, for input vector$x$, weight matrix$W$, and bias vector$b$ into probabilities. The output vector of the softmax layer is a vector$a$ with probabilities for each class.

Recurrent neural networks allow information produced by a neuron to be used as input together with inputs to the neural networks at the next time step \cite{f94}. Long Short-Term Memory (LSTM) neural networks are RNNs, with a complex structure combining various activation functions. LSTM include memory cells for keeping gradient information that can be fed back into the neuron’s activation functions \cite{hs97}. Various network topologies exist. For example, modifiable self-connections decide whether to overwrite a memory cell, retrieve it, or retain it for the next time step \cite{gs01}. LSTM are widely used for natural language processing and other tasks with time-variant data, even over long periods of time \cite{schmidhuber15}.

Convolutional neural networks (CNN) are variants of neural networks that specializes on multi-dimensional data\footnote{ CNN works on 2D matrices but can handle several of those matrices, called \textit{channels}. Therefore, it can handle, for instance, RGB images with one channel for each color. Standard ML models operate on 1D vectors; CNN operate on 2D matrices. With data streams and tensors this becomes multi-dimensional.} that supports processing of image datasets \cite{lecun2015deep}. Convolutional layers transform input data filter matrices (kernels). Low-level kernels detect simple entities, such as edges whereas higher order kernels are sensitive to more complex visual structures \cite{ksh17}. Thus, CNN filters and transforms data with the help of massive application of low-level mechanisms, such as max pooling, padding, and striding. Stacking layers of large amounts of neurons enables complex visual computing operations with high quality, such as object recognition and object tracking in videos.

\vspace{1\baselineskip}
\textbf{Unsupervised learning}

In unsupervised learning, teaching a model can be accomplished without ground truth data of dependent variables$y$. Hence, loss functions cannot be used for assessing model quality, but need to be replaced by heuristics or other quality metrics. The focus is only on direct inference of properties of the probability density function of dataset$X$ \cite{htf09}. From complex datasets, simpler approximation models are derived by using principal components models, multidimensional scaling, self-organizing maps, and principal curves. Other classes for extracting model abstractions are clustering and association rules.

Cluster models are centered around the concept of proximity; that is, the distance $d(x_{i},x_{y})$ between two instances of a dataset $X$. The overall distance $D$ with normalized weights $w_{k}$ for each dimension is generally formulated as follows: $D\left(x_{i},x_{j}\right) = \sum_{k = 1}^{p}w_{j}\ast d_{j}(x_{ik},x_{jk})$ \cite{htf09}. Distance function $d$ can be instantiated by the Minkowski distance with different settings for parameter $p$ ($ p = 1$: Manhattan distance, $p = 2$: Euclidean distance) or self-defined functions. Clusters are abstractions of probability density function $P(X)$ and mediate interpretation of datapoints by domain experts that goes beyond model properties. Unknown datapoints are directly associated with clusters and, thus, inherit cluster interpretations.

K-means is a simple model often used for clustering. It uses an iterative procedure that stop if a threshold $\varepsilon$ is undercut or if a maximum number of iterations is reached. For a given $k$, K-means determines the distance between all datapoint $x\in X$ and datapoint $k_{i}\in K\subseteq V$ with $V$ the p-dimensional vector space of $X$, $\left\vert K\right\vert  = k$ and associates label $k_{j}$ to a datapoint $x_{i}$ if distance $d(k_{j},x_{i})$ is the smallest for all $k\in K$. Then, for each label, $k_{j}$ is replaced by $k_{j}^{'}\in V$ that is the center of all datapoints $x_{i}$ with label$k_{j}$. This process is repeated until $\sum_{i,j}^{k}d\left(k_{i},k_{j}\right)<\varepsilon$ or a maximum number of iterations is reached. Datapoints $x_{i},x_{j}$ connected to a cluster $k_{s}$ ($ x_{i},x_{j}\in X(k_{s}))$ are stronger connected than with any $x_{m}\in X\left(k_{r}\right)$ with $s\neq r$.

Hierarchical clustering iteratively merges clusters based on a group distance metric. For classification tasks, loss functions are defined on the percentage of correct and incorrect classifications and how these change with different $k$ values.

\vspace{1\baselineskip}
\textbf{Generative models}

While standard approaches of unsupervised learning models attempt to find unknown labels for instances, generative models are used to learn underlying distributions of given data that can be used for sampling from these models; that is, the generation of novel instances indistinguishable from input data. For instance, a \textit{generative adversarial neural network} (GAN) \cite{g14} learns a generator’s distribution $p_{G}$ over data $x$ by a neural network $G\left(z\right),$ with$z$ taken from an input noise variable of $p_{z}\left(z\right),$ and uses a discriminator $D(x)$ neural network for making decisions as to whether the examples are real or fake. The goal of the generator is to generate artificial examples from the learned distribution that the discriminator cannot distinguish from real examples, i.e.$D\left(G\left(z\right)\right)\approx 1$ with$z$ random noise. Models$G$ and $D$ compete with each other while increasingly becoming better; that is, $G$ generates more realistic output and $D$ gets better at discriminating fake from real. The situation between $G$ and $D$ is modeled as a game-theoretic min-max game \cite{g14}:

\begin{center}
$ \min_{\theta_{G}}\max_{\theta_{D}} [E_{x\sim p_{data}}\log D_{\theta_{D}}\left(x\right) - E_{z\sim p(z)}\log\left(1 - D_{\theta_{D}}\left(G_{\theta_{G}}\left(z\right)\right)\right)]$
\end{center}

with samples from the distribution of the real world$p_{data}$. The discriminator tries to maximize correct classification of real versus generated data samples by training model weights$\theta_{D}$ of the discriminator neural network. On the other hand, the generator tries to minimize the success of the discriminator by training a model weights$\theta_{G}$ of the generator neural network so that$D_{\theta_{D}}\left(G_{\theta_{G}}\left(z\right)\right)$ becomes close to 1. GANs are used for generating 2D images \cite{g14}, 3D models \cite{bsm19}, music \cite{ycy17} and even support arithmetical operations \cite{rmc15}.

\textit{Autoencoders} consist of one neural network used for encoding input data $x$ into$z\left(x\right)$ and one that is used for decoding results of the encoder $\hat{x} = d(z)$. The goal is to encode only the information in$z$ that is critical for reconstructing $\hat{x}$ with the smallest error. The smaller $z\left(.\right),$ the less information used. This shows the resemblance to principal component analysis (PCA) while, unlike PCA, a non-linear mapping between compressed representation and the original representation can be achieved. Autoencoder architectures are effective for reduction of dimensionalities. Variational autoencoders (VAEs) \cite{kw13} are used for content generation by using distributions of latent space $z$, i.e. $p(z\vert x)$ instead of $z\left(x\right)$.\footnote{ Both VAEs and GANs result in generative models, although neither is a subset of the other. VAEs model the data generating distribution explicitly as an infinite mixture of Gaussians, whereas GANs do so implicitly. } Complex distributions are approximated by variational inference that defines a set of Gaussian distributions $\mathcal{N(}g\left(x\right),h(x))$, with $g$ mean and$h$ variance dependent on $x$. Mean and variance functions are optimized by minimizing the Kullback-Leibler divergence between the approximation and the target distribution $p(z\vert x)$.

\vspace{1\baselineskip}
\textbf{Reinforcement learning}

\begin{center}
\begin{figure}[H]
\centering
\includegraphics[width=8.78cm,height=3.43cm]{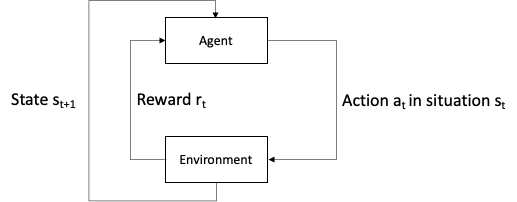}
\caption{Reinforcement learning.}
\label{fig:rl}
\end{figure}

\end{center}

Reinforcement learning combines machine learning with agent-based systems \cite{sb18}. A reinforcement learning model tries to maximize a reward function, such as winning a game (Silver et al. 2017) or a robot walking up a slope \cite{gs20}. The machine learning environment, including model and training mechanism, takes the role of an agent that predicts the best action according to an internal strategy without internal representation of the environment (model-free reinforcement learning). An action is performed in a specific situation and the agent receives a reward for this action (cf. Figure \ref{fig:rl}). This procedure ends when a final situation is reached. The goal of the agent is to maximize reward. The environment is an abstraction of the world in which an agent operates; that is, it is a model of the world. The world could be digital; for example, artificial, as in a chess game, or fully realistic and physical such as used for research on robotics. An agent learns by performing actions in an environment and receiving feedback in the form of rewards. By adjusting the model, the agent tries to find means for maximizing the total reward.

A reinforcement learning system is based on the concept of a Markov decision process. A state of a Markov decision process completely characterizes the state of the world under investigation (Markov property) by $\mathcal{(S,A,R,}\mathbb{P,}\gamma )$ with $\mathcal{S}$ the set of possible states, $\mathcal{A}$ set of possible actions, $\mathcal{R}$ distribution of reward, i.e. (state, action) pairs, $\mathbb{P}$ transition probability on actions in the environment and$\gamma$ discount factor. A policy$\Pi$ is a function from $\mathcal{S}$ to $\mathcal{A}$ that specifies which action is best to take in a given state. The objective is to find a policy$  \Pi^{\ast }$ that maximizes the cumulative discounted reward$\sum_{t>0}^{}\gamma^{t}r_{t}$ by taking a series of future actions.

In contrast to supervised learning with minimizing loss, reinforcement learning tries to find a sequence of actions and situations that maximizes the sum of rewards: $\Pi^{\ast } = arg\max_{\Pi }\mathbb{E}\left[\sum_{t\geq 0}^{}\gamma^{t}r_{t}\vert \Pi\right]$ with $s_{0}\sim p(s_{0}), a_{t}\sim \pi (.\vert s_{t}), s_{t + 1}\sim p(.\vert s_{t},a_{t})$. The value of a situation $s$ is calculated by the sum of rewards the agent expects under the given policy $\pi: V^{\Pi }\left(s\right)\mathbb{ = E}\left[\sum_{t\geq 0}^{}\gamma^{t}r_{t}\vert s_{0} = s,\pi\right]$ and the value of action $a$ performed in situation $s$ under a policy $\pi$ given by the Q-value function by accumulating the expected discounted reward $r: Q^{\Pi }\left(s,a\right)\mathbb{ = E}\left[\sum_{t\geq 0}^{}\gamma^{t}r_{t}\vert s_{0} = s,a_{0} = a,\pi\right]$. From this objective, the agent tries to find a policy that maximizes $Q: Q^{\ast }$. A standard approach optimizes $Q^{\ast }$ by satisfying the \textit{Bellman equation: }$Q^{\ast }\left(s,a\right) = r + \gamma \max_{a^{'}}(Q^{\ast }\left(s^{'},a^{'}\right))$ in any situation $s$ and using $Q^{\ast }$ for defining policy $\pi:  \pi\left(s\right) = \max_{a}(Q^{\ast }\left(s,a\right)).$ For problems with small action set $A$ and small state set $S$, $Q^{\ast }$ is defined by a look-up table, whereas, for complex environments, such as robotics, $Q^{\ast }$ is approximated by neural networks (deep q-learning): $Q(s,a,\theta )\approx Q^{\ast }(s,a)$ with parameters $\theta$. The difference to standard neural networks in supervised learning is that labels $y$ are unknown. The challenge is to use the Bellman equation with the current function approximation $Q(.)$ for calculating labels $y$; for example, Baron Munchausen getting himself out of the mud by pulling on his hair (Münchhausen trilemma)\footnote{ https://en.wikipedia.org/wiki/M$\%$C3$\%$BCnchhausen\_trilemma}. By using a squared loss function $L_{i}(\theta_{i})$, the difference between $y$ and the estimate of the neural network is used to determine the loss. The gradient of the loss (with holding $Q(.)$ fixed) is used for improving $Q(.)$ so that the next label $y$ is closer to $Q^{\ast }\left(.\right)$, assuming convexity.

\subsubsection{Model training}

Parametric machine learning models iteratively adjust model parameters so that the error on estimates for unknown input data is minimized. Examples include: linear regression, support vector machines, decision trees and neural networks.  This iterative adjustment process is called \textit{model training, }which searches through combinations of weights and model architectures, i.e. the configuration of components that will be fitted to data, such as the number, type, sequence and size of layers in a neural network. Model training compresses datasets into model parameters. In practice, model sizes can range from few bytes to over 100GB. GPT-3 \cite{b20}, for example, has approximately 325GB with 175B parameters of the language generation model represented with 16Bit precision floating point number \cite{d20}. The model fitting process creates a data generating function for generative tasks, that replicates output data similar to the unknown function underlying the data used for training. For classification, the function clearly discriminates. Fitting methods require at least as many training samples as there are parameters to be trained.\footnote{ This, at least, was the case for traditional machine learning, although recent research in neural networks suggests this may be false with shot learning and zero shot learning looking promising for some models.}

 

Several methods are used for optimizing model performance that involve either growing a model (e.g., gradient boosting) or adjusting weights of a fixed model (e.g., neural networks). In all cases, model optimization is guided by its associated objective function. Linear regression uses ordinary least square method. Maximum likelihood is a well-known method for finding the optimum values for the parameters by maximizing a negative log-likelihood function derived from the training data \cite{b06}. This is also used for logistic regression models. Decision trees and ensemble learning models, such as AdaBoost, use an additive expansion approach that adds weak learners for reduction of prediction errors \cite{htf09}. For neural networks, the loss function is optimized by using gradient descent where the gradient of the loss function on a dataset at hand is calculated with  \textit{backpropagation }\cite{htf09}. With gradient descent, each weight is slightly adjusted along the negative gradient according to its contribution to the result to minimize the loss function.

\vspace{1\baselineskip}
\textbf{Loss function}

\begin{center}
\begin{figure}[H]
\centering
\includegraphics[width=12.11cm,height=5.7cm]{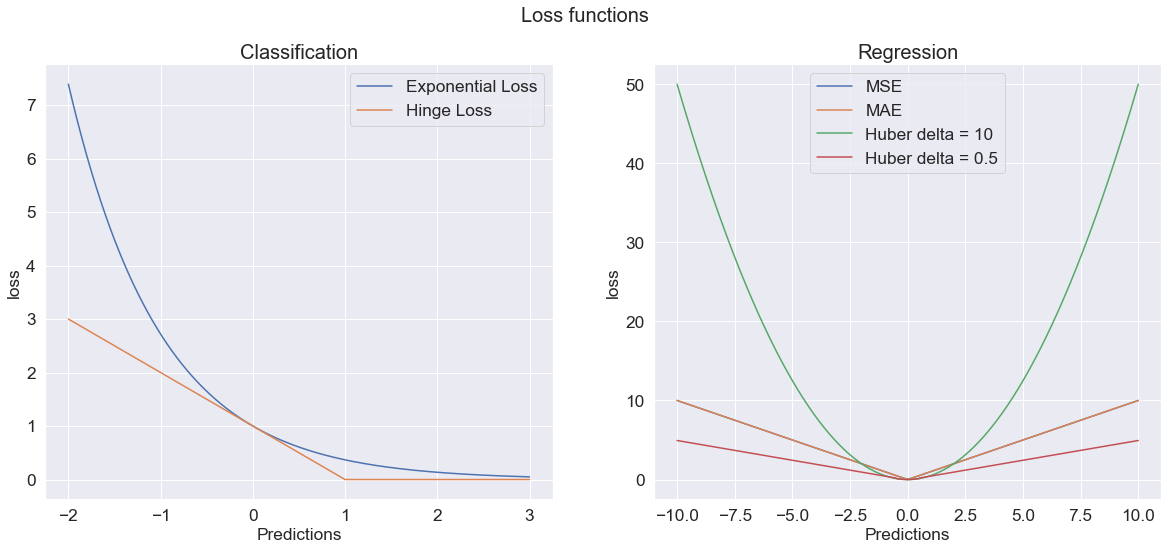}
\caption{Example loss functions for classification and regression.}
\label{fig:lossfct}
\end{figure}

\end{center}

Beside mean squared loss error (MSE), several other loss functions exist for regression tasks, such as mean absolute error (MAE) and a combination of both (cf. Figure \ref{fig:lossfct}), which is called the Huber loss \cite{h65}. For binary and multiclass classification, cross entropy and corresponding Kullback-Leibler divergence are used that assess the difference in entropy of training and testing data and the entropy of predictions (figure \ref{fig:lossfct}). Other loss functions are commonly used, such as hinge loss or exponential loss \cite{htf09}.

%
The goal of model training is to find a model parameter that minimizes the error of the selected loss function$ L(f)$:

\begin{equation}
\min_{w\in W}\sum_{i\in 1\ldots n}^{}L(f\left(w,x\right),y_{i})
\end{equation}

\vspace{1\baselineskip}
\textbf{Gradient descent procedure }

Instead of progressing through the entire search space of weights $W$, gradient descent is used for finding local minima for a given weight vector $w$ by calculating partial derivatives of $L$ on $w$. Because $L$ is to be minimized, the negative of the derivatives is used. For updating the weight vector $W$, a scalar step size $\eta$ is used that determines the size of adjustment. If stepsize $\eta$ is set too large, the risk of destabilizing the optimization procedure increases, whereas a step size that is too small assumes the risk of slowing down training and reaching the maximum number of iterations too early, before reaching the minimum (For details cf. section 4.3 in \cite{g16}). This equation shows the dependency of the training algorithm on the definition of the loss function.

$$w:=w - \eta \nabla_wL(f(w,x), y)$$


Function $f(w,x)$ contains the logic for computing an estimate $\hat{y}$. The difference between $\hat{y}$ and $y$ is the core of a loss function in supervised learning, because the loss is dependent on $f\left(.\right),$ which, in turn, is dependent on parameters $w$ (or$\theta$). Partial derivatives of $L(.)$ on $w$ adjust weights in the direction of a minimum of loss function $L(.)$.

\vspace{1\baselineskip}
\textbf{Model overfitting, bias and variance}

\begin{center}
 \begin{figure}[H]
\centering
\includegraphics[width=16.29cm,height=4.78cm]{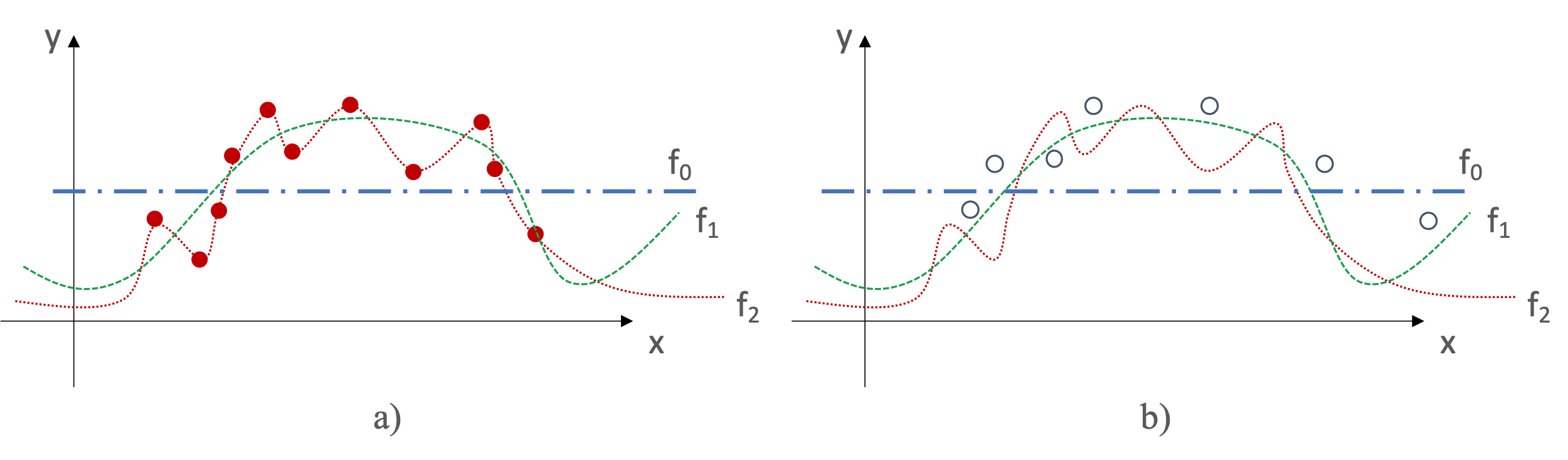}
\caption{Bias and variance; circles in b) indicate unseen data used for testing the models.}
\label{fig:bias}
\end{figure}

\end{center}

Not all machine learning models have the same capacity for capturing the signals embedded in a dataset. Linear models can only capture linear functions, whereas neural networks generally capture non-linear functions. However, learning algorithms that can produce models that can learn arbitrary relationships between inputs and outputs, so they might adapt to idiosyncratic data and outliers, and hence not generalize to new data. This trade-off is characterized by \textit{bias} and \textit{variance}. A model that is too simple for capturing the complexity of a function underlying a dataset has high bias (cf. linear function $f_{0}$ in Figure \ref{fig:bias}a)); that is, on average, it shows high error. A trivial model (cf. function $f_{2}$ in Figure \ref{fig:bias}a)) shows no error and has a bias of 0. Function $f_{1}$ is in the middle between $f_{0}$ and $f_{2}$ with respect to bias. When applied to unseen data (Figure \ref{fig:bias}b)), function $f_{2}$ shows a large error whereas function $f_{1}$ is better. However, function $f_{0}$ is not able to estimate new data well. The degree by which models perform worse on testing data than on training data is called the \textit{variance of a model}. Function $f_{2}$ shows low bias, but high variance. Function $f_{0}$ has high bias and low variance (because it performs poorly on training and testing data) whereas $f_{1}$ has low bias and lower variance than $f_{2}$. Function $f_{1}$generalizes better than $f_{2}$ because it has a lower loss on new data not present in the training data. Alternatively, $f_{1}$ is suffering less from \textit{overfitting} to training data than $f_{2}$. Function $f_{0}$ is \textit{underfitting} the data due to high bias and, thus, is not being specific enough to capture the underlying function. In practice, the greater the complexity of a model, the greater the tendency to overfit. This is accounted for by adding a regularization function $J(.)$ to the risk function $R(.)$ that adds a penalty to more complex models. This is shown in Figure \ref{fig:bias}.

\begin{center}
 \begin{figure}[H]
\centering
\includegraphics[width=7.93cm,height=3.18cm]{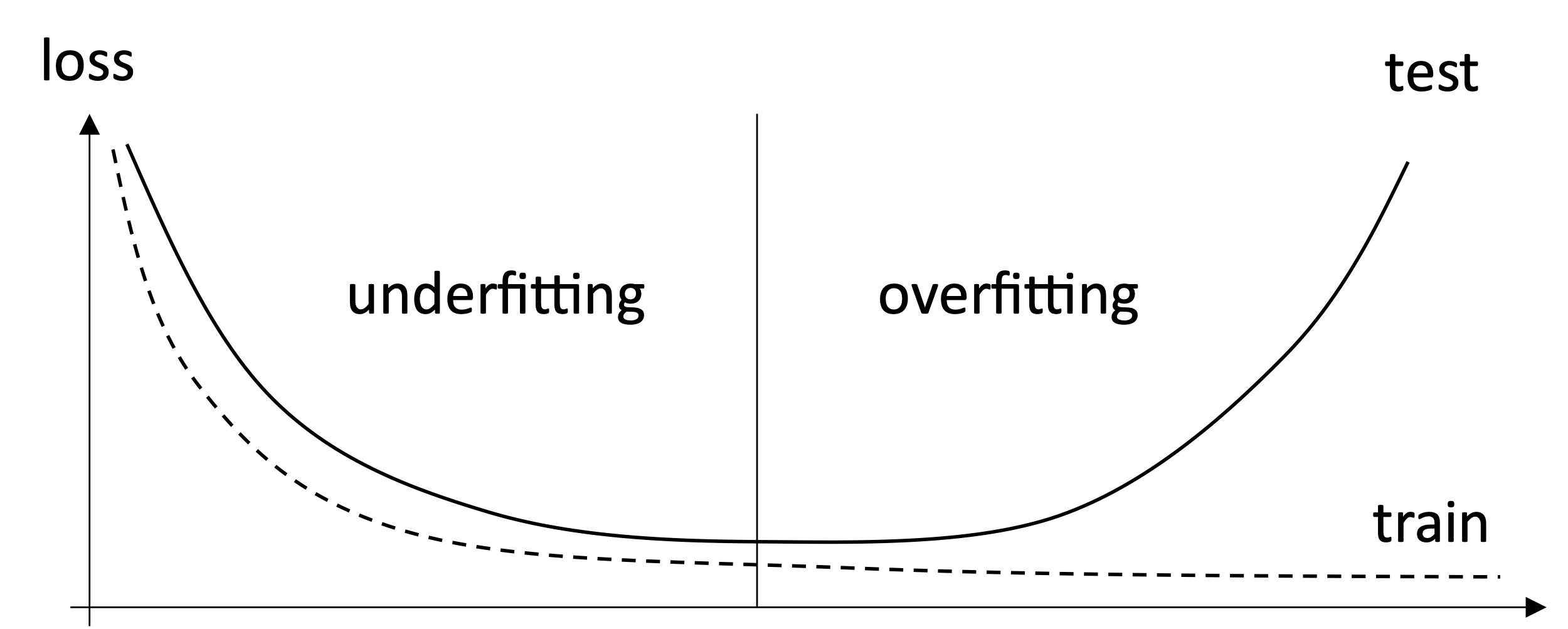}
\caption{Overfitting and underfitting.}
\label{fig:overfitting}
\end{figure}

\end{center}

In general, model search trees is a process to optimize the search for a model that minimizes the loss on training and unknown testing data via analyzing underfitting and overfitting behavior (cf. Figure \ref{fig:overfitting}).

%
%
\subsection{Data Science Development Cycle}

The development of information systems based on machine learning is still progressing. Major platform providers have published their own processes, such as Google’s \textit{Train-Evaluate-Tune-Deploy} workflow\footnote{ https://cloud.google.com/ai-platform/docs/ml-solutions-overview} or Amazon’s \textit{Build-Train-Deploy} model \footnote{ https://aws.amazon.com/getting-started/hands-on/build-train-deploy-machine-learning-model-sagemaker}. By analyzing five development models, including CRISP-DM (cross-industry standard process for data mining) \cite{wh00}, we identify six phases in the machine learning development cycle, as shown in Table \ref{tab:mldevproc} \cite{km06}. Proposed models should map business requirements into data requirements. Technically, data is prepared according to data requirements and processed with appropriate data mining technologies. Deployment mainly consists of presenting discovered knowledge. Research in information systems has traditionally adopted data mining processes which focus more on the variables under investigation and less on technologies.


\begin{table}[]
	\begin{adjustbox}{max width=\textwidth}
	\begin{tabular}{|p{3.42cm}|p{3.32cm}|p{3.0cm}|p{3.25cm}|p{3.28cm}|} 
		\hline
		\textbf{Generic model Kurgan and Musilek} \cite{km06} & \textbf{Shmueli and Koppius }\cite{sk11} &\textbf{Chambers and Dinsmore }\cite{cd14}  & \textbf{Goodfellow et al.} \cite{g16} &  \textbf{Data science development process}\\ \hline
		Application domain understanding& Goal definition & Defines business needs &Determination of goals  & Problem understanding \\ \hline
		Data understanding& Data collection and study design & \multirow{2}{3.0cm}{Build analysis data set} &  Establish a working end-to-end pipeline&  Data collection\\ \cline{1-2}\cline{5-5}
		\multirow{2}{3.42cm}{Data preparation and identification of data mining technology}&  Data preparation&  &  & Data engineering \\ \cline{2-2}
		& Exploratory data analysis &  &  &  \\ \cline{2-2}
		& Choice of variables &  &  &  \\ \hline
		Data mining& Choice of potential methods &  \multirow{2}{3.25cm}{Build predictive model} & Instrument the system well to determine bottlenecks in performance & Model training \\ \cline{1-2}\cline{4-5}
		Evaluation& Evaluation, validation and model selection &  &  Repeatedly make incremental changes such as gathering new data, adjusting hyperparameters, or changing algorithms&  Model optimization\\ \hline
		Knowledge consolidation and deployment& Model use and reporting & Deploy predictive model & \textit{Beyond discussion} &  \textit{Model Integration }\\ \hline
		\textit{Beyond discussion}&  \textit{Beyond discussion}&  \textit{Beyond discussion}&  \textit{Beyond discussion}& Analytical decision making \\ \hline
	\end{tabular}
\end{adjustbox}
	\caption{Data science development processes.}
	\label{tab:mldevproc}
\end{table}

\vspace{15\baselineskip}
Research on deep learning adds development processes that focus on data processing pipelines because ML models grow excessively in size, sophistication, and training costs. Therefore, focusing on performance issues, identifying bottlenecks, and optimizing hyperparameters, are critically important for deep learning models (cf. Table \ref{tab:mldevproc}) \cite{g16}.

The goal of a data science project is to \textit{approximate an unknown function} by a fitted statistical model that exhibits an estimated function, which generates results (predictions) by obeying domain and technical constraints, while maximizing performance goals. A learner abstracts from all possible learning models and makes hypotheses on the unknown functions that relate input data with results (in a hypotheses space) \cite{d12}. This means that data scientists choose data (aka features) and data representations. They also select model candidates that are hypothetically capable of finding a fitted function,$f$, showing satisficing performance. This step requires explicit representations for data, functions, constraints, and performance goals including objective functions that can be scrutinized as part of subsequent analysis and explanations of results. It is especially important that the objective function used for assessing the quality of a trained model is, not only defined based on its technical purposes, but also supports domain requirements. For example, if domain experts are interested in identifying all features with an impact on the results, this would be contradicted by using a L1/Lasso norm that tends to eliminate features with small impacts and, thus, favor a sparse function$f$. Domain experts and data scientists must agree on project goals, generally, and on the level of complexity of the trained model, specifically, in accordance with the problem statement.

By integrating the different views, we now propose a \textit{data science development process model} that consists of the phases identified in Figure \ref{fig:mldev}.

\begin{center}
{\large    \begin{figure}[H]
\centering
\includegraphics[width=12.68cm,height=11.24cm]{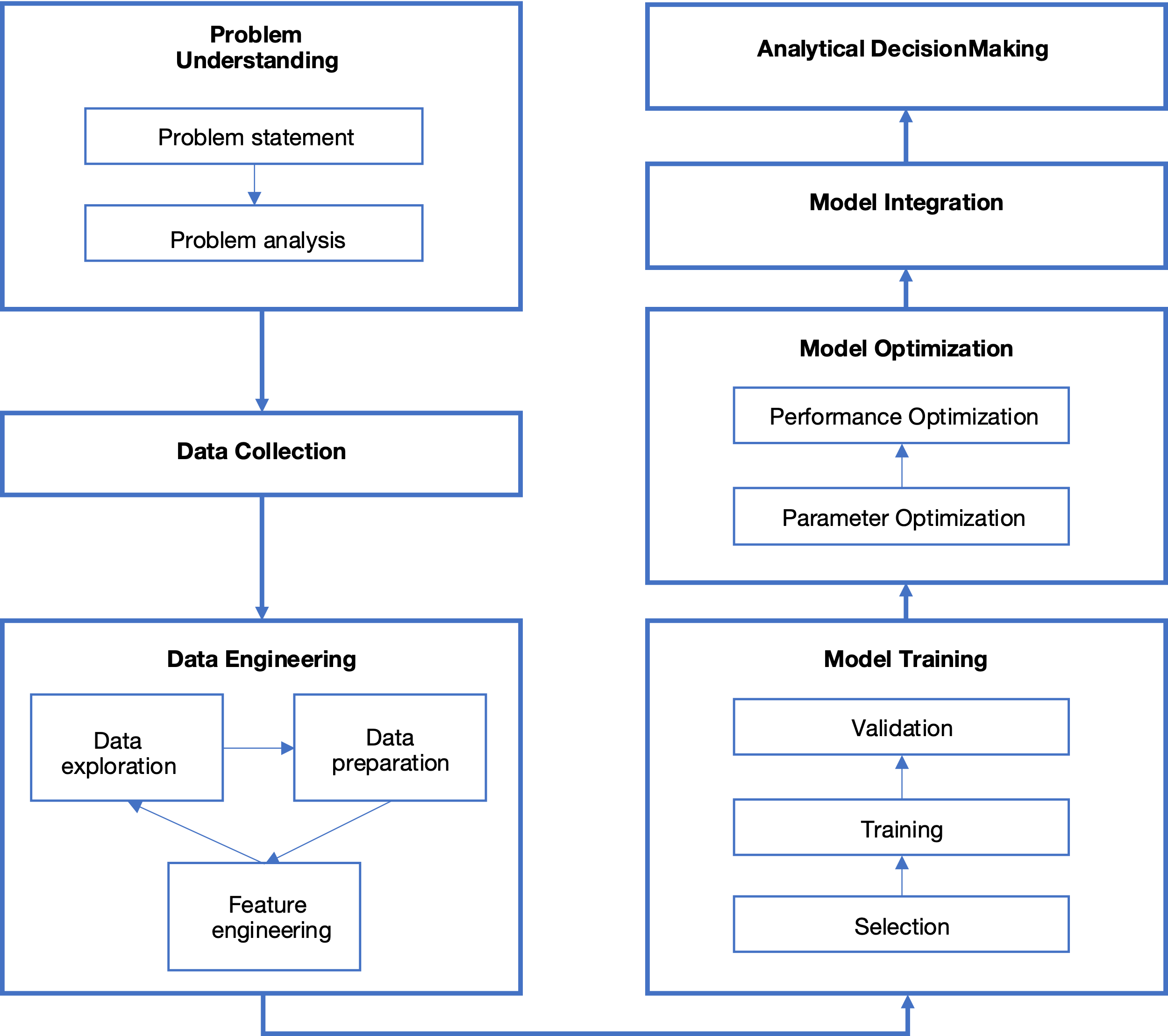}
\caption{Data science development process.}
\label{fig:mldev}
\end{figure}
}
\end{center}


\subsubsection{Problem understanding}

Complex environments require managers to make decisions under increasing uncertainty. By digitalizing many common processes, business environments have been able to adopt machine learning technologies, such as manufacturing \cite{w16}, finance \cite{rundo19}, marketing \cite{hr21} and also social media \cite{s20}. Data science is applied to decision problems that can be addressed by statistics and machine learning technologies. Data science projects translate problem statements provided by domain experts into project definitions for data scientists. Given a problem statement, a data scientist tries to find solutions to a decision problem by posing three questions: (1) is there a mathematical formalization for this decision problem and a solution path based on linear algebra and statistics, and, if so, (2) is this solution path implementable on some software platforms and (3) is this implementation scalable for production?

\textit{Problem understanding} starts with a problem statement and description of the data science project by domain experts. The results from both are discussed with data scientists until a shared understanding exists. Software engineering has many examples that support the importance of shared understanding \cite{a00}. Even though software engineering and conceptual modeling have investigated means for properly building shared knowledge, little has diffused into machine learning research. A strong emphasizes on data alone diminishes the importance of domain knowledge and the role that domain experts play in designing ML-based information systems.

A \textit{problem statement} is a hypothesis, created by a domain expert, which asserts that a decision problem can be solved by a computational process. A data scientist analyzes a problem statement for gaining a proper understanding of the problem. The problem understanding phase is highly iterative. Typically, neither the problem statements nor the data scientist’s understanding of the domain is sufficient. Conceptual modeling provides a rich toolbox for supporting shared understanding between domains and technical experts. A problem statement describes decision making situations and parameters that influence decision making. Uncertainties and external influences might influence the decision-making process. Because decision making is embedded into a business context, performance goals, such as key performance indicators (KPIs) or response time behavior of decision processes are defined \cite{g17}.

During \textit{problem analysis} it is important to assess whether a problem statement can be translated into a data science problem that is feasible to solve, given desired performance goals. The problem analysis phase also includes project management issues, such as negotiation and definition of human, data and computational resources, milestones, and time plans. During problem analysis, data scientists start investigating whether the problem can be understood as a classification or regression task and whether this becomes accessible by supervised, unsupervised or reinforcement learning approaches. For any data science project, it is crucial to determine the accessibility of data, data size, and data quality. Care is required if data needs to be collected. In general, the resources required for data collection are grossly underestimated, but have a direct impact on data quality.

\subsubsection{Data collection}

Data is the core object in data science projects. Data is not just collected by some business processes, but also by sophisticated means, such as the Internet of Things (IoT) sensors, remote sensing, social media, financial markets, weather data, supply chains, and so forth. In this sense,  data becomes an economic asset (data product \cite{wsf95}) exchanged via data ecosystems \cite{oj19}.

Data collection is constrained by data requirements derived by a problem statement and problem analysis, specification of data sources and corresponding data types, and volume, and quality requirements. Data with sufficient quality is a precondition for quality results of data science projects. Data quality is described using four main categories: (1) \textit{intrinsic} including \textit{accuracy}, (2) \textit{contextual} including \textit{relevancy} and \textit{completeness}, (3) \textit{representational} including \textit{interpretability}, and (4) \textit{accessibility }\cite{ws96}. Additionally, some researchers have added \textit{availability} as another main category \cite{cz15}. Data quality strategies are distinguished as: (1) \textit{data-driven} and (2) \textit{process-driven}. A data-driven strategy improves data quality by data modification. A process-driven strategy tries to modify the process by which data is collected \cite{b09}. Numerous data quality methods exist that span steps for evaluation of costs, assignments, improvement solutions, and monitoring, with numerous quality metrics \cite{b09}.

\subsubsection{Data engineering}

After data has been made available, it is cleansed, explored, and curated. For univariate data these procedures overlap with data mining (cf. CRISP-DM) \cite{wh00}. This process is more demanding for multi-variate data, such as image and video data with multi-dimensional features with channels (e.g., for RGB colors). Time series data, such as that provided by sensors, often contain missing data that needs to be replaced by meaningful data fitting with a temporal context \cite{l18}. More sophisticated exploration and preparation methods are used for unstructured data, such as texts and auditory data. Auditory data is usually transformed into textual data from which core structures are extracted by text mining, including keyword selection and linguistic preprocessing (e.g., part-of-speech tagging, word sense disambiguation) \cite{hnp05}. An example from health care is found in Palacio and Lopez (2018) \cite{p18}.

Data exploration is used to understand a dataset in detail with respect to the domain. Descriptive statistical analysis provides standard metrics, such as mean, standard deviation of variance of single features, and correlation values between two features, whose values must remain within the boundaries of the application domain. In addition to statistical analysis, semantic analysis exploits domain requirements for assessing data validity.   Ontological representations \cite{g15} may be associated with features to support domain experts in understanding the datasets. The domain can also restrict constraints on data values and the range of acceptable values. For instance, the concept $``$blood pressure$"$ has an associated constraint that blood pressure values cannot be negative. Thus, domain requirements enable domain experts to understand data sets and assess their quality and, in this way, reflect some of the semantics of the real world. Domain requirements can be simple statements, such as feature ranges, or complex conceptual models with cascades of requirements that need to be tested carefully. For some domains, theories with formal representations exist.

Datasets are rarely collected in highly controlled laboratory environments but, instead, collected in different environments under dynamically changing conditions. Datasets are mixed, merged, and added with features that are not necessary for the data science task at hand. Feature engineering provides methods for identifying the features that are relevant and those that are not. However, this task is highly dependent on both the domain and problem statement. Technically, feature engineering is domain-specific and requires intuition, creativity, and $``$black art$"$ \cite{d12}. Technical feature engineering alone can result in negative side-effects if, for example, features are dropped that are relevant or merged by incorrect means. Relevant features subsequently increase performance of the trained model \cite{bl97}.

Large univariate and multivariate datasets are generally difficult to analyze at an item level. However, flawed results of data science projects are often caused by missing an understanding of the structure and meaning of a dataset. Research in statistics has developed standard visualizations of probabilistic data, such as visualization of density functions plus visualization of statistical measures, such as box plots, histograms, scatter plots, normal Q-Q plots and quasi-visualization, such as correlation matrix and confusion matrix. Legendary are Hans Rosling’s data visualizations that make transparent what is hidden in raw data.\footnote{ https://www.ted.com/playlists/474/the\_best\_hans\_rosling\_talks\_yo} Exploration of multivariate data is much more complex. For instance, analyzing whether images in an animal data set actually show buildings requires either many people \cite{e07} or models that have been developed on other datasets.

When a domain expert and data scientist have a common understanding of the data set and single features, data is prepared for analytical processing. Data preparation includes: (1) \textit{data exploration}, (2) \textit{data preparation}, and (3) \textit{feature scaling}. Similar to ETL (extract-transform-load) in data mining, the \textit{data exploration} phase processes and transforms raw input data into a dataset of sufficient quality\textit{. Data preparation} includes statistical procedures for handling missing data, data cleansing, and data transformation by normalization, standardization, and reduction of dimensions (e.g., Principal Component Analysis (PCA)). Data preparation is a $``$black art$"$ that needs to be transformed into a $``$white art$"$. This goal of data interpretability aligns with the need for interpretability of models and explainable AI (XAI) for making $``$black boxes$"$ of models transparent to users (e.g., \cite{rsg16}).

Data interpretability requires the data preparation steps to obey data constraints as part of domain constraints. For example, missing data deals with replacing unknown entries by computed values, such as a mean or median. Most frequently, $``$not a number$"$ (NAN) is used or samples are deleted that have missing values. Data constraints for features guide data scientists in selecting appropriate procedures. Social sciences have experience with handling outliers according to different outlier categories: error outliers, interesting outliers, and influential outlier \cite{agj13}. When lacking proper domain understanding, outliers are often deleted in data science projects because of negative effects on performance measures. However, interesting, and influential, outliers are anomalies relative to the dataset that potentially provide insights. For example, the fundamental purpose of the ATLAS (A Toroidal LHC ApparatuS) project on finding the Higgs boson \cite{a08} focused on finding anomalies, so deleting outliers would have rendered this project useless. Therefore, data constraints define the limits on what is theoretically possible in a data science project.

Data constraints are derived from domain theories. They describe ranges, rules, and invariants as well as functions on features. For instance, ranges constraint meaningful feature values, whereas rules describe dependencies between features of a sample. For example, if age is x years (today), then the date of birth is within a specified range of years$[today – x – 1, today – x]$ . Invariants are strong assertions that hold within features (for instance, feature on gender are required to be evenly balanced) or between features (for instance,$birthday < day of death$). In addition to textual descriptions, ranges, rules and invariants can be formally modeled using various formalisms, such as subsets of predicate logic \cite{mhs09}, constraint logic programming \cite{jm94}, constraint satisfaction formalisms \cite{v89}, and constraint formalisms for object models \cite{rg98}. More challenging is ensuring that data constraints are valid when data transformation is applied.

Validation data annotation ensures that data preparation obeys domain constraints and generates datasets that are meaningful at both the feature level and the dataset level. The basis for semantic data preparation includes four categories for data quality: accuracy, relevancy, representation, and accessibility \cite{wsf95, ws96}.

Missing data is a major concern in almost any data preparation phase. Various imputation strategies \cite{v18} are applied for replacing missing data by random values, mean or median values, or most frequent value; using feature similarity in nearest neighbor models; removing features; or applying machine learning approaches, such as DataWig \cite{b19}.\ \ Current machine learning models only work with numerical values. Therefore, categorical or textual data is transformed into numerical representations. A standard technique for categorical data is \textit{one-hot-encoding,} which adds binary features for each category. Preprocessed textual data is often categorical and is either mapped onto numerical indexes or transformed by one-hot-encoding.

Beyond standard imputation and encoding, data is transformed in various ways. Integration of features can lead to more expressive, additional features. For instance, if one feature is income and another is number of people per household, adding a feature that divides income by number of people per household can provide valuable information for predicting educational development. Several machine learning models, as well as gradient descent, use differences between features, such as KNN, k-means and SVM. Therefore, features with larger scales have more influence than smaller scales. Normalization (min-max scaling) and standardization are standard feature scaling procedures that tend to improve model training and prediction quality.

Normalization is applied if the data does not follow Gaussian distribution. Empirically, models that do not presuppose specific distributions, such as KNN, Perceptrons \cite{mp17} and neural networks, can improve prediction performance by normalized data. Similar improvements can be achieved by standardizing data for use in distribution-dependent models.

Feature engineering is selects, transforms, adds, constructs, or replaces features in such a way that it improves model training and model performance, without changing feature semantics. Data scientist’s prior knowledge and skills are needed for organizing data representations so that discriminative information become accessible \cite{bcv13}. Various methods are used for creating additional features from input features, such as calculating differences, ratios, powers, logarithms, and square roots \cite{heaton16}. For text classification, correlation-based methods are used, such as information gain \cite{yp97}. Semantic similarities of concepts derived by using ontologies \cite{p07} are used for feature ranking and feature selection (e.g., \cite{p07}).

Representation learning is a current research topic that attempts to automatically extract representations of data, such as posteriori distribution of some explanatory factors underlying observed input. These factors decrease the complexity of feature engineering because they can be used as guidance or even as input to supervised learning models \cite{bcv13}.

\subsubsection{Model training}

Model training includes \textit{selection}, \textit{training} and \textit{evaluation} of models. Training a model means adjusting the model parameters to the data. Based on an objective function. Supervised learning models adjust weights according to loss gradients for minimizing, for instance, the sum of squares for regression and minimizing cross-entropy for classification \cite{htf09}. Unsupervised learning models use the sum of distances, and reinforcement learning use updates based on reward evaluations. In the early phases of data science projects, it is usually not clear which machine learning model will exhibit the best performance. Therefore, several model types with hyperparameter ranges are often tested against each other. This iterative exploration phase narrows down prime candidates for subsequent phases.

With the introduction of many different types of model architectures within a short period of time (due to the surge in the popularity of machine learning), guidelines and modeling patterns are increasingly important. To date, the architecture designs of machine learning models emerge from the practical needs of machine learning experts. This knowledge slowly diffuses to less experienced designers. Conceptual modeling, with its capabilities for abstracting the real world, can help make machine learning architectures more accessible and practically useful \cite{st16}. Elements of a model driven architecture (MDA), including UML, provide languages for describing machine learning model architectures. MDA could aid in the selection of implementation of algorithms, based on users’ requirements. Furthermore, object-oriented design patterns \cite{g95} provides a basis for technical design patterns for constructing machine learning model designs.

A general challenge for designing model architectures lies in the appeal of complex models. Even unexperienced machine learning architecture designers are inclined to prefer recent and more complex models over older and simpler models. Model design requirements are necessary that constraint minimum and maximum complexities of model designs according to problem statements and associated goal models and goal constraints. At an abstract level, model design requirements describe guidelines \cite{st16}. At a technical level, model design requirements provide information on required capabilities of model units on various levels. For instance, there could be requirements on the capability of neuron types (e.g., plain neuron or LSTM neuron) or pattern of connections between neurons; e.g., fully connected, or filters for convolutions are on the lowest level. Larger structures of layered neurons are called a topology of a network \cite{m11}). Intermediate requirements encompass the number of layers, building blocks of layers (e.g., LSTM layer, softmax layer) and general mechanisms, such as attention. Top-level requirements describe the model design space. For example, it might require the use of linear models only or models for which theoretical guarantees exist, such as those associated with complexity classes or optimality criteria.

Depending upon the datasets used, model designs have a major impact on model performance. Making requirements on performance ranges explicit will further restrict the model design space. Using performance requirements at design time is either based on heuristics or is probabilistic because of the unknown function underlying the dataset, making the actual model performance unknown. Means for expressing heuristics on the relationships between performance requirements, datasets, and model types include heuristic rules, constraints, and logical expressions. Relationships can also be learned, given enough data on performance, models, and datasets. This multi-dependency between dataset, model design, and performance requirements carry knowledge that is important for any model designer and decision maker. The more experience that is accessible, the better a data scientist can select model designs that fulfill targeted performance ranges. A small, clarifying example for this argument is a neural network with two input features (x1, x2), one fully connected hidden layer with just two neurons, and an output layer with one neuron for adding activations. Given data from two classes that are embedded in one another (i.e., cannot be separated by a line), this model will probably not exhibit high performance (i.e., small loss) with respect to accuracy because model complexity is not specific enough and, thus, underfits the dataset.\footnote{Example, cf. https://t1p.de/oc2w} A performance requirement is expressed as a loss on misclassified samples of less than 10$\%$, probabilistic knowledge on performance ranges for this small neural network, and a binary classification task with 500 datapoints will inform model designers about a likely mismatch between the modeling task and performance requirements at design time. In this case, performance requirements are achieved by adding another neuron to the hidden layer.

Structural dependencies in model design tasks have an impact on resources spend on model training, parameter optimization and, subsequently energy and time consumption. So far, this knowledge is part of a data scientist’s $``$black art$"$. Making this crucial knowledge explicit by conceptual model representations is important for managing data science projects and businesses. Constraints languages, such as OCL (Object Constraint Language) \cite{rg98}, are means for describing and evaluating requirements between dataset, model design and performance requirements at model design time and, subsequently, for model evaluation when actual model performance is assessable. Model evaluation tests whether actual performance fulfills performance requirements.

Assessing the ability of a model to generalize is important for performing well on unseen data. The more complex a model, the better it can adjust to training data (low bias), although it might overfit and work less well on unseen testing data (high variance) \cite{htf09}. The goal is finding a model and a model architecture with a minimum of absolute training and testing loss and a minimum distance between them.

Data sets are split into several parts used for training, validation, and testing. Splitting data is often based on heuristics, for instance, 50$\%$ training, 25$\%$ validation, and 25$\%$ testing. The training set is used to train as many models as there are different combinations of model hyperparameters. These models are then evaluated on the validation set, and the model with the best performance (e.g., the smallest loss or highest accuracy) on this validation set is selected as the final model. This model is retrained on training and validation data with the selected hyperparameters. Then, model performance is estimated using the test set. It is assumed that the model generalizes well if the validation error is similar to the testing error. Finally, the model is trained on the full data.

If datasets are small, training and validation is carried out with the same dataset. Folded \textit{cross-validation} and \textit{bootstrapping} are used for iterative assessment of model accuracy. Cross-validation separates the training dataset into partitions (folds) of the same size. One fold is separated for assessing model performance and the others used for training. Average performance is determined by repeating this process with all folds. Bootstrapping draws samples from the training dataset with replacement and trains a model for a specified number of times. Accuracy is assessed by averaging over all iterations \cite{htf09}.

\subsubsection{Model optimization}

With unlimited resources, machine learning models can train and evaluate to find an optimal system configuration. Business analytics and data science, as well as related research has continued to require attention \cite{ccs12, m18}. Model complexity increases excessively, making a brute-force approach infeasible. Optimization tasks are ubiquitous in machine learning. A key optimization task is finding weights that minimize loss in supervised learning or finding policies that support a goal best in reinforcement learning. Gradient descent and stochastic gradient descent are basic algorithm for weight optimization. However, more efficient algorithms are used in practice that adds a momentum vector for speeding up gradient updates (e.g., Adam optimizer \cite{kb14}).

Most machine learning tasks are controlled by external parameters, called \textit{hyperparameters} that constrain a model’s search space; for instance, the value of k in KNN, maximum depth or number of trees in random forest, or number of layers and neurons per layer in neural networks. With brute-force, k would be the range of all positive integers and a threshold for performance. Experience in the domain and previous experiments might have shown that k$=$$\{$4,$\ldots$,7$\}$ are most likely candidates for minimizing the loss function and optimal accuracy. Therefore, experience indicates that k$=$$\{$1,2,3$\}$ is not worth training. For real-valued hyperparameters, this problem become even more pressing. Grid search is a greedy procedure for finding the best hyperparameter settings by using all hyperparameter combinations. This only works for small datasets and small number of hyperparameter combinations. Several approaches exist for automatic hyperparameter optimization \cite{h11}. Recently, AutoML systems have been introduced, such as Auto-sklearn \cite{f20} or AutoKeras \cite{jsh19}, that provide automatic optimization across hyperparameter settings with an emphasis on neural networks.

Configuration parameters of relational databases barely make it into scientific discussions. The difference is that hyperparameters directly affect finding at least locally optimal models. Setting ranges for hyperparameters too large might result in excessive resource requirements whereas too small ranges might threaten the search for the best model. Requirements on hyperparameter are influenced by the domain, dataset, expertise and previous modeling tasks, but, most of all, by the model type and its implementation. Similar to performance requirements, hyperparameter requirements are an open field for conceptual modeling. Hyperparameter requirements can simply set parameter ranges. Alternatively, hyperparameter requirements can describe the complex dependencies between business requirements, goals, resource models, performance requirements, services requirements, and others. With enough knowledge captured by hyperparameter requirements, companies can optimize their resources by invested in a ML-based service development that can result in shorter time-to-market of products and services.

The \textit{performance} of a model is assessed by analyzing results of predictions for testing data. For classification, the number of items that are correctly and incorrectly classified are analyzed. For a binary case, four cases are differentiated. Two are correct (positives and negatives are correctly classified); two cases make opposite predictions (false negatives, false positives). A confusion matrix separates these four cases: true positive (TP), true negatives (TN), false positives (FP) and false negatives (FN). Sensitivity, $\frac{TP}{TP + FN}$, is a measure for positive cases and specificity, $\frac{TN}{TN + FP}$, for negative cases. If false negatives and false positives are rare, sensitivity and specificity are close to 1. In practice, it depends on the domain and the decision task as to which metric is most important. For instance, in healthcare, there is stronger emphasis on sensitivity. Alternatively, precision $\frac{TP}{TP + FP}$, and recall, $\frac{TP}{TP + FN}$, are used with recall the same as sensitivity and precision the percentage of correct true cases modified by false positives. The F1-score combines precision and recall in one metric which is useful in cases with no clear preference for precision or recall. Finally, accuracy in binary classification is the percentage of true classifications over all samples, $\frac{TP + TN}{TP + FP + TN + FN}$.

The loss function for classification uses cross-entropy:$H\left(p,q\right) =  - \sum_{k = 1}^{K}p_{k}\ast \log\left(q_{k}\right)$ with$p$ probability of ground truth and$q$ probability of predicted categories. If probability$q$ is close to probability$p$, cross entropy will be close to 1. The difference between cross-entropy$H\left(p,q\right)$ and the entropy of probability$p$, i.e.$H(p)$, is called Kullback-Leibler Divergence$D_{KL}(p\vert\left\vert q\right) = \sum_{k = 1}^{K}p_{k}\ast \log\left(\frac{q_{k}}{p_{k}}\right) = H\left(p,q\right) - H(p)$.$D_{KL}$is used as a metric for the performance of a classification model.

For a regression task, proportion of declared variance$R^{2} =\frac{\sum_{i = 1}^{n}\left(\hat{y_{i}} -\overline{y}\right)^{2}}{\sum_{i = 1}^{n}\left(y_{i} -\overline{y}\right)^{2}}$ is often used that is close to 1 if residuals between ground truth$y_{i}$ and estimates $\hat{y_{i}}$ are small. Because loss functions for regression tasks normally use squared residuals, weights are adjusted for minimizing residuals and, therefore,$R^{2}$.

After training a machine learning model based on a risk function, model performance is evaluated by performance metrics. The performance metric value for a model requires domain-dependent interpretation. For instance, \textit{sensitivity} and \textit{specificity} results for a binary classification on healthcare diagnosis typically favors sensitivity (percentage of true positives $=$ $``$ill$"$), over specificity (percentage of true negatives $=$ $``$not ill$"$). Relative performance values are used for model comparison whereas absolute performance values determine whether a model is good enough. Thus, performance metrics are operationalizations of quality requirements that a model needs to satisfy; that is, model performance expressed by a performance metric is required to exceed a quality threshold. Conceptual modeling can support performance evaluation in two ways: 1) selection of performance metrics; and 2) threshold for absolute performance for selected performance metrics. Performance requirements constraining the selection of performance metrics depend on the domain and the modeling task. For instance, for classification tasks healthcare domain prefers sensitivity/specificity over precision/recall.

Performance thresholds are target of extensive debates in research domains (for instance, discussion on threshold for confirmatory factor analysis, CFA \cite{n67})\textsuperscript{ }and, thus, carry deep knowledge. In the simplest form, requirements on performance thresholds are single numbers but can be expanded to intervals and distributions (similar to confidence intervals). From a scientific point of view, performance requirements must be defined before model training so that performance results of model evaluation on testing data can be assessed without bias. In practical applications, performance results are input for decision makers for making decisions on project progress and future business. If performance results seem promising by getting closer to performance requirements, positive decisions on investments in subsequent development phases are more likely. However, performance requirements are not absolute but, rather, adapt to developments in a particular field. For instance, NLP (Natural Language Processing) adopts performance metrics from computer vision (e.g. Intersection-Over-Union, IOU \cite{e10}) but also define new performance metrics, such as comprehensiveness and sufficiency within the context of explainable NLP \cite{d19}.\textsuperscript{ }Performance requirements could be a large area of research with descriptions of performance requirements needed. Goal models are required for mediating business goals and performance results. Specification languages are needed for properly representing, communicating, and eventually automatically reasoning on performance representations.

\subsubsection{Model Integration and Evaluation}

ML-based information systems in decision making are recognized as an important topic for both research and practice. Many applications use machine learning model types that can be directly evaluated by humans. For legal and business reasons, ML-based information systems are required to explain their results by means accessible by non-technical domain experts (explainable AI - XAI). Linear regression models, logistic regression models, decision trees and support vector machines can be all scrutinized; much more effort is required for complex ensemble models, such as those based on XGBoost. Single predictions by deep learning models and reinforcement learning models are based on myriads of simple, highly interconnected calculations that make direct understanding by domain experts impossible. For instance, a decision to stop a production line due to a ML-based prediction requires strong arguments and explanations. A recent approach involves fitting simpler surrogate models close to local areas of a prediction and using \textit{surrogate models} for explanation, such as Individual Conditional Expectation (ICE) plots \cite{goldstein15}, Local Interpretable Model-agnostic Explanations (LIME)\ \ \ \ \cite{rsg16} and Shapley Additive Explanations (SHAP) \cite{ll17}.

\subsubsection{Analytical Decision Making}

\begin{center}
\begin{figure}[H]
\centering
\includegraphics[width=13.12cm,height=5.95cm]{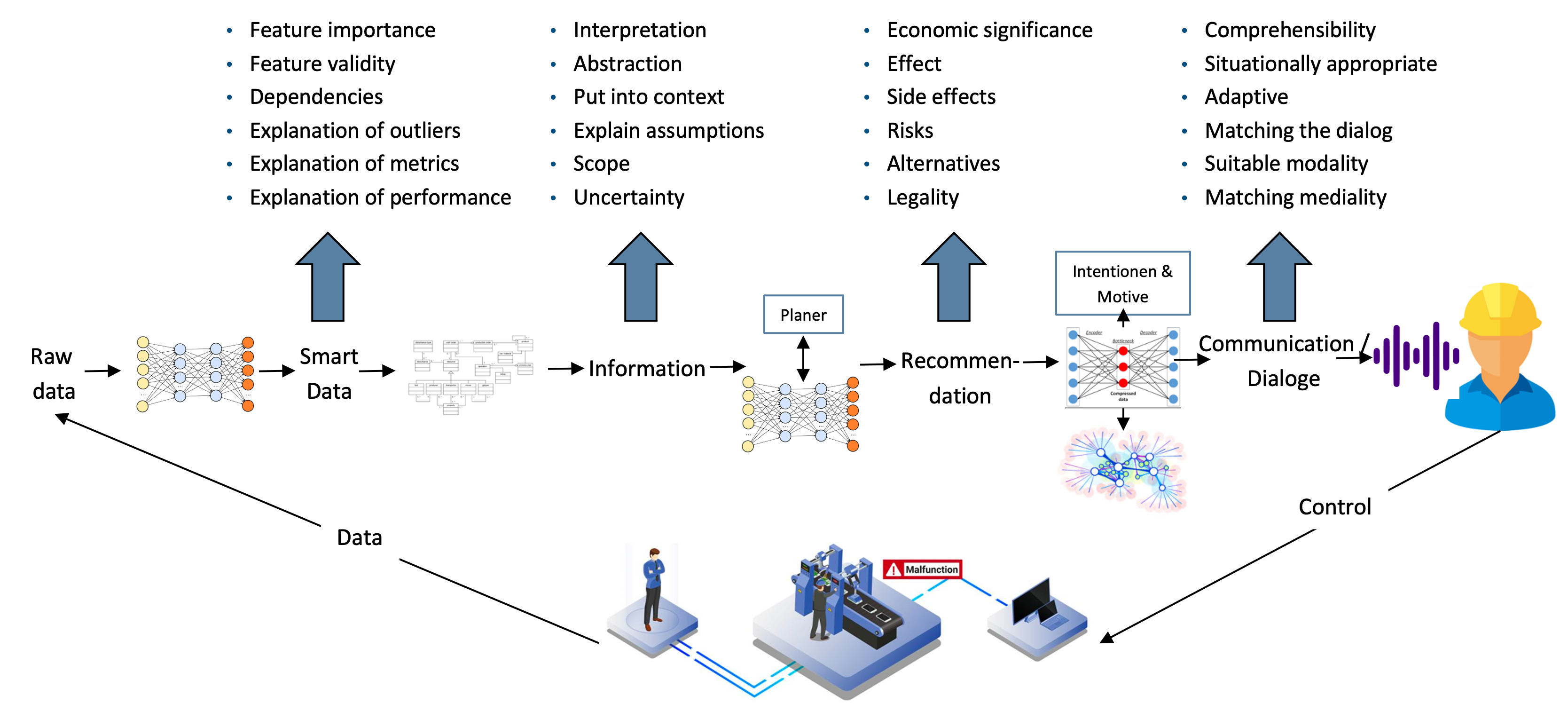}
\caption{Explainability framework.}
\label{fig:xai}
\end{figure}

\end{center}

Decision makers need more than just explanations for predictions. The performance of models strongly depends on data, so decision makers must: scrutinize raw data; pre-processed smart data \cite{ss16}; identify the semantic information used for merging and processing data; identify the objectives of the data scientists who developed machine learning models; and estimate economic effects, side effects, risks, and alternatives. Decision makers also need to understand potential semantic losses (\textit{lost in translation}). Examples of these requirements are included in the Explainability Framework in Figure \ref{fig:xai}.

Figure \ref{fig:xai} captures the important concept of \textit{Explainable AI} (XAI), which shows how raw data is operated on to progress to information that can be used to make recommendations to a user \cite{a20}. Users need to understand the explanation of how the output is obtained. This enables the user to consider the explanation and assess whether it is necessary to rework a problem.

\section{Conceptual Modeling for Machine Learning}

Practical machine learning models are only useful within a given domain, such as games, business decisions, healthcare, politics, or education. When embedded into information systems, machine learning models must follow laws, regulations, societal values, morals, and ethics of the domain, and obey requirements derived from business objectives. This highlights the need for conceptual modeling to address the $``$black box$"$ challenge of complex machine learning models. Conceptual models help transform business ideas into structured, and sometimes even formal, representations that can be used as precise guidelines for software development. Therefore, they help structure the thought processes of domain experts and software engineers, for building a shared understanding between these groups and for providing languages by which information system implementations can be understood, scrutinized, revised, and improved \cite{msk11, css10}. Conceptual modeling of machine learning can also make it easier to gain skills by abstracting machine learning technologies with the help of model-driven software engineering and automatic code generation \cite{bucchiarone20}.

Complex machine learning models, such as deep learning models, are often considered as $``$black-boxes,$"$ which are not well-scrutinized. Machine learning experts and data scientist perceive domain knowledge as a quarry from which ideas and initial guidelines can be extracted. ($``$We begin by training a supervised learning (SL) policy network$p_{\sigma }$ directly from expert human moves.$"$ \cite{shm16}). The mechanistic nature of reinforcement learning requires exploration of any changes in environments \cite{iyy02} and, thus, is focused on how, rather than why, the decision-making process occurs \cite{b14}.\textsuperscript{ }In domains, such as video games, the basis on which a decision has been made is not important. However, in business domains, decision making requires: \textit{trust} in recommendations; sufficient \textit{understanding} of the reasoning processes and the underlying assumptions behind a recommendation; \textit{legal} and \textit{ethical} obligation adherence; and support of \textit{stakeholder} requirements. As extreme examples, a ML-based system could recommend laying off all employees or investing in weapons for the extermination of mankind. No serious decision maker will follow such recommendations without scrutinization. However, the decision maker will ask for explanations; look at the data used for training; ask for second opinions and recommendations of alternative models; analyze software development procedures and requirement documents; talk to software engineers and data scientists; hire external experts for unbiased views; and probably much more. This requires documentation and identification of the representations that were used for designing and building this ML-based information system, and understanding how they will help to explain a system’s behavior and recommendations.

There are differences between machine learning systems and, for instance, database systems. Database systems implement domain knowledge that has been adopted by domain experts. In contrast, ML-based information systems are not intended to implement prior knowledge, but rather, find useful patterns for making predictions given input data. These patterns may lead to theoretically interesting questions that could guide subsequent research, as is typical in biomedical research. For domains, such as gaming, designing, research and development, music and art, this freedom for finding innovative patterns and making unprecedented predictions is appreciated. For domains, such as legal decision making, production and manufacturing, healthcare, driving, operating chemical and power plants and military, highly reliable and trustworthy information systems are required that follow laws, ethics and values. Thus, conceptual modeling methods and tools should enable users to scrutinize, understand, communicate, and guide the entire lifecycle of ML-based information systems. This motivates the need for a general framework that aligns design, development, deployment, and usage of ML-based information systems for decision making.

\subsection{Framework for conceptual modeling in data science}

The alignment between business strategy and business operations with its IT strategy and IT operations is an enabler for competitive advantages \cite{m04}. Conceptual models provide specification languages for capturing business requirements that can be translated into software requirements \cite{u11, pastor2011}. They include primitive terms, structuring mechanisms, primitive operations, and integrity rules \cite{mcy99}, with the entity-relationship model representative of a semantic specification language \cite{c76}. Dynamic sequences of activities are captured by process models, such as event-driven process chains, UML activity diagrams or BPBM models. For domain experts, requirement models and specifications are generally too abstract, so  early requirements analysis attempts to capture stakeholders’ intentions \cite{ckm02} and goals \cite{yu97}. Conceptual models are translated and refined by software-oriented requirements languages until they can be used as a basis for implementation.

Alignment of ML-based information system development with business goals and strategies is at an early stage of research and understanding \cite{lps18, m20}. Therefore, guidelines and frameworks are needed to identify the research topics that need to be studied and to support the progression of the needed research. For example, ML technologies are used to explore potential cost reduction (e.g., via predictive maintenance), but used less for business innovations. Decision makers might be reluctant to employ machine learning because of possible poor data quality and black-boxed ML algorithms \cite{cc20}. Although the entity-relationship model was initially introduced to gain a $``$unified view of data$"$ \cite{c76}, it has, after many decades of research, been extended to business goals, intentions, processes and domain ontologies \cite{stl15,f16,emp03}. A domain ontology provides a set of terms and their meaning within an application domain \cite{g93}, with many domain ontologies having been created and applied; however, there are many quality assessment challenges \cite{mss18,ms19,b05}.

In response to the need to align machine learning and business, as well as the challenges in doing so, Table \ref{tab:framework} provides a framework for incorporating conceptual models into data science projects.

\ \ \ \


\begin{table}[]
	\begin{adjustbox}{max width=\textwidth}
		\begin{tabular}{|p{4.23cm}|p{3.75cm}|p{7.75cm}|} 
			\hline
			\textbf{Data Science Main Phases} & \textbf{Sub-Phases} &\textbf{Conceptual Modeling Concepts, Methods and Tools}  \\ \hline
			\multirow{2}{4.23cm}{Problem understanding}&Problem statement & Business requirements, goal model\\ \cline{2-3}
			& Problem analysis & Business requirements, goal model; legal and ethical requirements; Data requirements \\ \hline
			Data collection& &Data requirements; Data quality; legal and ethical requirements, business requirements \\ \hline
			\multirow{3}{4.23cm}{Data engineering}& Data exploration& Data requirements; Legal requirements
\\ \cline{2-3}
			& Data preparation& Ontologies; Domain models
\\ \cline{2-3}
			& Feature engineering& Ontologies; Domain models\\ \hline
			\multirow{3}{4.23cm}{Model training}&Selection& Business requirements, legal requirements, performance requirements and conditions for acceptance \\ \cline{2-3}
			& Training& Performance requirements \\ \cline{2-3}
			& Validation& Performance requirements\\ \hline
			\multirow{3}{4.23cm}{Model optimization}& Parameter optimization& Domain model, resilience requirements\\ \cline{2-3}
			& Performance Optimization&Domain model, resilience requirements \\ \hline
			Model integration& & \multirow{2}{7.75cm}{Business requirements;Goal model; Legal requirements; Ethical requirements; Data requirements}
\\ \cline{1-2}
			Analytical decision making& & \\ \hline
		\end{tabular}
	\end{adjustbox}
	\caption{Framework for incorporating conceptual models into data science projects.}
	\label{tab:framework}
\end{table}

\textbf{Problem understanding}

Using Machine learning models within a business context requires providing solutions to business problems. Research in innovation distinguishes between $``$technology-push$"$ and $``$need-pull$"$ \cite{s67}. From the technology-push perspective, adoption of machine learning is the driving force for competitive advantages \cite{pm85}. This view is challenged by the long sequence of failure that AI has suffered over many decades. For instance, Google’s Duplex dialog system that impersonates a human, raises ethical concerns and affects trust in businesses, products and services \cite{o19}. Other examples use machine learning for visual surveillance, which breaches privacy laws or uses machine learning on social media data to influencing political debates. Legal \cite{buiten19} and ethical requirements \cite{by14} increasingly influence design decisions on ML-based services. This resolves uncertainties for data scientists who are challenged by unclear ethical and regulatory requirements \cite{vb19}.

Generally, elicitation of business needs and business requirements within the context of ML-based information systems is a novel field of research. However, it is dominated more by questions and challenges, than answers, such as the lack of domain knowledge, undeclared consumers, and unclear problem and scope \cite{bvc19}.  Proposed approaches to business intelligence have a strong overlap with ML-based information systems \cite{h19}. Because the classes of ML-based information systems are broader than business intelligence systems, a wider range of stakeholders need to provide input to problem understanding. Qualitative methods are often used to  understand business needs and elicit requirements \cite{mv12, mj11}. In general, conceptual modeling provides a large set of modeling approaches that help heterogenous teams gain a shared an understanding of the strategic and operational business needs and goals, as well as the constraints associated with ML-based information systems. This includes shared understanding on performance and quality requirements for data, models, and predictions.

Goal modeling may become a key contribution of conceptual modeling to ML-based system development \cite{l19}. Nalchigar et al. (2021), for example,\ \ propose three views for modeling goals for ML-based system development: business view, analytics design view, data preparation view \cite{nyk21}.

\vspace{1\baselineskip}
\textbf{Data collection}

Any information system depends on input data.  ML-based information systems even extract behavior from data, which is why data is so important. Database and information systems emphasize the importance of data schema, with web-based open data increasingly annotated with semantic markers (e.g., Gene Ontology, YAGO, dpPedia, schema.org). In specific contexts, data standards are available for different industries (e.g., eCl@ss or UNSPSC). For streaming data, as prevalent in real-time systems and the Internet of Things applications, new standards are defined, such as OPC-UA for processing semantically annotated data in distributed environments and using machine learning based on, for instance, MLlib with Apache Spark.

Appropriation of existing data sources falls into two categories: open data and proprietary data. Many governments maintain open data repositories, such as Data.Gov in the USA and Canada, and GovData in Germany. Wikipedia extracts are provided by dpPedia (dbpedia.org). Access to proprietary data depends on contractual agreements because raw data is generally not protected by copyright laws, whereas audio and image data can claim copyright protection if deemed to be artwork. Work on digital rights management (DRM) has developed proprietary and open solutions for protecting media data, such as music and videos and enforcing license management \cite{v06}. Application of DRM on operational data, such as data by the Internet of Things systems, requires analytical run-time environments that implements DRM standards \cite{liang18}. Blockchain approaches enforce immutable exchange of data and execution of contractual obligations \cite{xx19}. Large Internet companies follow a business model that centralizes data via cloud infrastructures and provides access to data via market mechanisms, such as auctioning. Alternatively, federated data platforms favor decentralized data repositories that are connected via data exchange protocols (e.g., GAIA-X in Europe).

Data collection depends on data requirements \cite{vb19} that provide a precise understanding of the type of data and data quality necessary for finding an application. Entity-relationship models and its derivatives are proper means to represent the requirements for data collection. These models can be used for storing, screening and interpreting of data collections in the sense of ETL-processes of data mining \cite{ss17}. Data integration from multiple sources with heterogenous data schema requires ontology-based matching and mapping \cite{es07}. Data requirements for univariate data overlaps with modeling approaches. Multivariate, graph-oriented, and textual data require research on extended modeling mechanisms. Beside alignment with business requirements, data requirements also capture crucial legal and ethical requirements. Examples include constraints on the origin of the data, as well as its quality.

Data quality has a major influence on model performance and, thus, the utility of a ML-based information system. Consistency and completeness are two major indicators of data quality \cite{kls14, plw02} with further research on data quality needed for the adoption of external data sources. Additionally, recent developments in data ecosystems, such as GAIA-X, shows the importance of modeling legal and contractual requirements \cite{c20}.

\vspace{1\baselineskip}
\textbf{Data engineering}

Data requirements provide a basis from which to consider data transformations.  \cite{jm15}. Data requirements capture \textit{semantical,} \textit{structural} and \textit{contextual} descriptions. Semantical descriptions represent information about data types, and their accepted interpretations; e.g., pressure is record in Pascal. Structural descriptions provide constraints on the form of the data; e.g., sample rate ranges, acceptable percentage of missing values, and accuracy of sensors used for collecting data. Structural descriptions are related to data quality, and also capture descriptive information, such as the time and location of data capture Contextual descriptions represent the constraints of a domain and the context within which a ML model is intended to be used, which includes requirements preventing biases or demanding coverage. Thus, data models capture semantical, structural, and contextual descriptions and provide information about obtaining data requirements.

Besides requirements, data engineering also extracts knowledge about data that has not been visible previously. The number of dimensions of a data space can be reduced (e.g., by principal component analysis) or additional dimensions added (e.g., by one-hot encoding of categorical dimensions). Combining dimensions requires theoretical understanding (e.g., of physical mechanics when combining mass and acceleration into force and velocity into energy). Dependencies between data and machine learning models require data engineering. For instance, models that use gradient descent work best if data dimensions are first standardized and normalized for multiplication.

Data requirements represent dependencies between data dimensions for constraining data transformations. Results for data explorations and data transformations are fed back into enhanced data requirements for capturing additional semantics. For instance, a typical first step in data engineering is correlation analysis between input data that is visualized for data scientists, but lost afterwards. Because data exploration, data preparation, and feature engineering generate rich knowledge about data, enhanced data requirements become important. Domain experts can scrutinize this knowledge about the data before it is used for model training.

\vspace{1\baselineskip}
\textbf{Model training}

Training complex machine learning model is resource-intensive and strains computation, energy and financial resources. Therefore, the declaration of functional and non-functional requirements guides model training and provides boundaries. Regulations and legal rules put requirements on data and model behavior, energy consumption, and sustainability. To date, research on legal requirements mainly focuses on the behavior of a machine learning model with respect to interpretability and explainability, especially as a consequence of European laws and  the General Data Protection  Regulation (GDPR) \cite{bibal20}. However, for commercial settings, the selection of machine learning models is tedious due to the need to avoid potential intellectual property infringements. It, thus, requires in depth technological and legal analysis, both before and after model selection. For example, decision makers might want to avoid spending extensive resources on training ML models, to later realize that they have violated license infringements. As models become more complex and stacked on top of each other, legal descriptions become even more important.

After model training, various descriptions characterize functional and non-functional model behaviors, including model performance. Conceptual modeling practices have a long tradition of capturing such characterization in concise conceptual models. These models can then be used to extend or combine ML and integrate them into information systems.

\vspace{1\baselineskip}
\textbf{Model optimization}

After model training, model optimization fine tunes the model parameters to ensure performance requirements are achieved. Doing so, requires updating conceptual models associated with the ML models. Resilience is a meta-requirement that describes a system’s capability under disturbances, such as lower data quality or fewer parallel processes capabilities, than expected. A resilient machine learning system does not deny service under disturbances, but gracefully degrades. At the end of model optimization, all requirements and corresponding system documentations must be reviewed and updated.

\vspace{1\baselineskip}
\textbf{Model integration}

Model integration resolves technical problems by addressing functional and non-functional requirements. This phase overlaps with traditional system integration that includes requirements for repair enablement, transparency, flexibility, and performance \cite{h00}. Requirements for the final analytical decision phase met business, legal and ethical requirements.

\vspace{1\baselineskip}
\textbf{Analytical decision making}

Integrated ML models that fulfil model and data requirements should support business requirements as represented by conceptual models including goal requirements. Interpretability is important for any business decision making system. The entire stack of conceptual models, fully integrated with data and ML models, provide an important source for interpretability. Shallow integration only provides approximate estimates of system behavior. Full integration requires provable guarantees. Both the conceptual model stack and guarantee mechanisms require further research.

The liability for recommendations made by a ML-based information system is common in any service-oriented business. Legislators and scholars have started to demand higher levels of transparency and explainability of AI and ML technologies \cite{b19}. Technical solutions for explainable AI (XAI) (cf. section  3.2.6) are initial attempts that need to be aligned with legal and regulatory requirements.

Table \ref{tab:spec} provides examples of specification languages known in conceptual modeling. Proven specification approaches exist for business, functional, and non-functional requirements. Nomos is used for legal requirements \cite{s12}. Data requirements resemble those for database systems and linked data. Specification approaches for ethical requirements, machine learning models, performance requirements, interpretability, and resilience also require further development and refinement.


\begin{table}[]
	\begin{adjustbox}{max width=\textwidth}
		\begin{tabular}{|p{2.25cm}|p{9.23cm}|p{4.5cm}|} 
			\hline
			\textbf{Topic} &	\textbf{Definition}	&\textbf{Example specification languages}\\ \hline
			Business &	description of business process that are related to the strategy and the rationale of on organization&	I*, BPMB, UML, BIM, URN/GRL \cite{a10}, BMM \cite{OMG}, DSML \cite{gmp09}

\\ \hline
			Legal&	Goals that choices made during the ML development are compliant with the law (based on \cite{s09})
&
			Nomos, Legal GRL \cite{gar14}

\\ \hline
			Ethical \cite{by14}
&
			Compliance with principles, such as transparency, justice and fairness, non-maleficence, responsibility and privacy \cite{jiv19}.
&
			textual
\\ \hline
			Data \cite{vb19}
&
			Requirements on semantics, quantity and quality of data	&ER, UML, RDF, OWL, UFO, OCL
\\ \hline
			ML Model \cite{jm15}
&
			selection of architectural elements, their interactions, and the constraints on those elements and their interactions necessary to provide a framework in which to satisfy the requirements and serve as a basis for the design \cite{pw92}.
&
			Finite state processes, labeled transition systems \cite{v03}

\\ \hline
			Functional &	statements of services the system should provide, how the system should react to particular inputs, and how the system should behave in particular situations. \cite{s04}
&
			BPMN, UML, EPC, KAOS, DSML \cite{f13} \cite{gmp09}

\\ \hline
			Non-functional	& A non-functional requirement is an attribute of a constraint on a system \cite{glinz2007non}
&
			UML, KAOS
\\ \hline
			Performance &	expressed as the quantitative part of a requirement to indicate how well each product function is expected to be accomplished \cite{c12}
&
			Rules quantified by metrics
\\ \hline
			Interpretability &	Interpretable systems are explainable if their operations can be understood by humans. \cite{ab18}
&
			Qualitative rules
\\ \hline
			Resilience&	ML models that gracefully degrade in performance under the influence of disturbances and resource limitations &	Rules quantified by metrics
\\ \hline
		\end{tabular}
	\end{adjustbox}
	\caption{Specification languages.}
	\label{tab:spec}
\end{table}

\subsection{Example}

The framework for incorporating conceptual models into data science projects (Table \ref{tab:framework}) is illustrated by the following example.

\vspace{1\baselineskip}
\textbf{Problem understanding}

The objective is to predict whether a female person has diabetes, recognizing that diabetes is a widespread disease that is difficult to manage.  The problem is addressed based on a dataset\footnote{ https://www.kaggle.com/uciml/pima-indians-diabetes-database} from the society of Pima Native Americans near Phoenix Arizona collected by the US National Institute of Diabetes and Digestive and Kidney Diseases. The Pima are a group of Native Americans living in central and southern Arizona and in Mexico in the states Sonora and Chihuahua. In the US, they live mainly on two reservations: the Gila River Indian Community (GRIC) and the Salt River Pima-Maricopa Indian Community (SRPMIC). The GRIC is a sovereign tribe residing on more than 550,000 acres with six districts. They are involved in various economic development enterprises that provide entertainment and recreation: three gaming casinos, associated golf courses, a luxury resort, and a western-themed amusement park.

Two SRPMIC communities, Keli Akimel O'odham and the Onk Akimel O'odham, have various environmentally based health issues related to the decline of their traditional economy and farming. They have the highest prevalence of type 2 diabetes in the world, leading to hypotheses that  diabetes is  the result of: genetic predisposition \cite{wfh84}, a sudden shift in diet during the last century from traditional agricultural crops to processed foods, and a decline in physical activity. In comparison, the genetically similar O'odham in Mexico have only a slighter higher prevalence of type 2 diabetes than non-O'odham Mexicans.

The Pima population of this study has been under continuous study since 1965 by the National Institute of Diabetes and Digestive and Kidney Diseases because of its high incidence rate of diabetes. Each community resident over 5 years of age has been asked to undergo a standardized examination every two years, which includes an oral glucose tolerance test. Diabetes was diagnosed according to World Health Organization Criteria; that is, if the 2 hour post-load plasma glucose was at least 200 mg/dl (11.1 mmol/l) on any survey examination or if the Indian Health Service Hospital serving the community found a glucose concentration of at least 200 mg/dl during the course of routine medical care \cite{k78}.

In a study by Smith et al (1988) \cite{s88}, eight variables were chosen to form the basis for forecasting the onset of diabetes within five years in Pima Indian women. Those variables have been found to be significant risk factors for diabetes among Pimas or other populations \cite{k78}:

\begin{enumerate}
	\item Number of times pregnant

	\item Plasma Glucose Concentration at 2 Hours in an Oral Glucose Tolerance Test (GTIT)

	\item Diastolic Blood Pressure (mm Hg)

	\item Triceps Skin Fold Thickness (mm)

	\item 2-Hour Serum Insulin Uh/ml)

	\item Body Mass Index (Weight in kg / (Height in in))

	\item Diabetes Pedigree Function

	\item Age (years)$\#$

	\item Outcome (diabetes: binary)

\end{enumerate}
The criteria applied were as follows.

\begin{itemize}
	\item The subject was female.

	\item The subject was$\geq$ 21 year of age at the time of the index examination.

	\item Only one examination was selected per subject. That examination was one that revealed a nondiabetic Glucose Tolerance Test (GTIT) and met one of two criteria: 1) diabetes was diagnosed within five years of the examination; or 2) a GTIT performed five or more years later, failed to reveal diabetes mellitus.

	\item If diabetes occurred within one year of an examination, that examination was excluded from the study to remove those cases that were potentially easier to forecast from the forecasting model. In 75$\%$ of the excluded examinations, diabetes mellitus was diagnosed within six months.

\end{itemize}
The goal of the project is to develop a machine learning model that predicts diabetes with high accuracy. Business requirements, business goals or performance goals are not given. Full privacy needs to be guaranteed according to HIPAA privacy rule.\footnote{ https://www.hhs.gov/hipaa/index.html} This dataset is problematic for privacy reasons because it relates to an identified tribe. Results of the analysis are associated with the tribe and can lead to discrimination. Publication of results would, most likely, require consent by the Pima people.

Development of data science solutions is generally conducted by multi-disciplinary teams consisting at least of domain experts and data scientists but usually includes software developers and functional experts, such as marketing, sales, product development and finance. Overcoming barriers given by technical complexities of machine learning and software engineering, modeling goals is an important means for shared understanding \cite{msk11}. Conceptual modeling (as opposed to problem modeling) introduces various types of goals between actors: functional goals, non-functional goals \cite{v01, y97} and soft goals \cite{mcn92}, both of which can be useful for this application.  Functional goals of ML-based information systems are similar to procedural information systems while non-functional goals refer to expected system qualities and help to align understanding and work of all team members. Important goals for the development of machine learning solutions include: (1) data quality as key indicator for data engineering results; (2) accuracy and performance as key indicator for model training and model optimization; and (3) runtime behavior as key indicator for model integration and analytical decision making. Goal models can help to synchronize the work of both actors and even increase creativity \cite{hml15} by clearly stating goals, events, dependencies and required resources as a means for overcoming barriers in development projects that leverage machine learning technologies. Soft goals explicate goals for the work relationship between actors.

\begin{figure}[H]
\includegraphics[width=13.73cm,height=6.96cm]{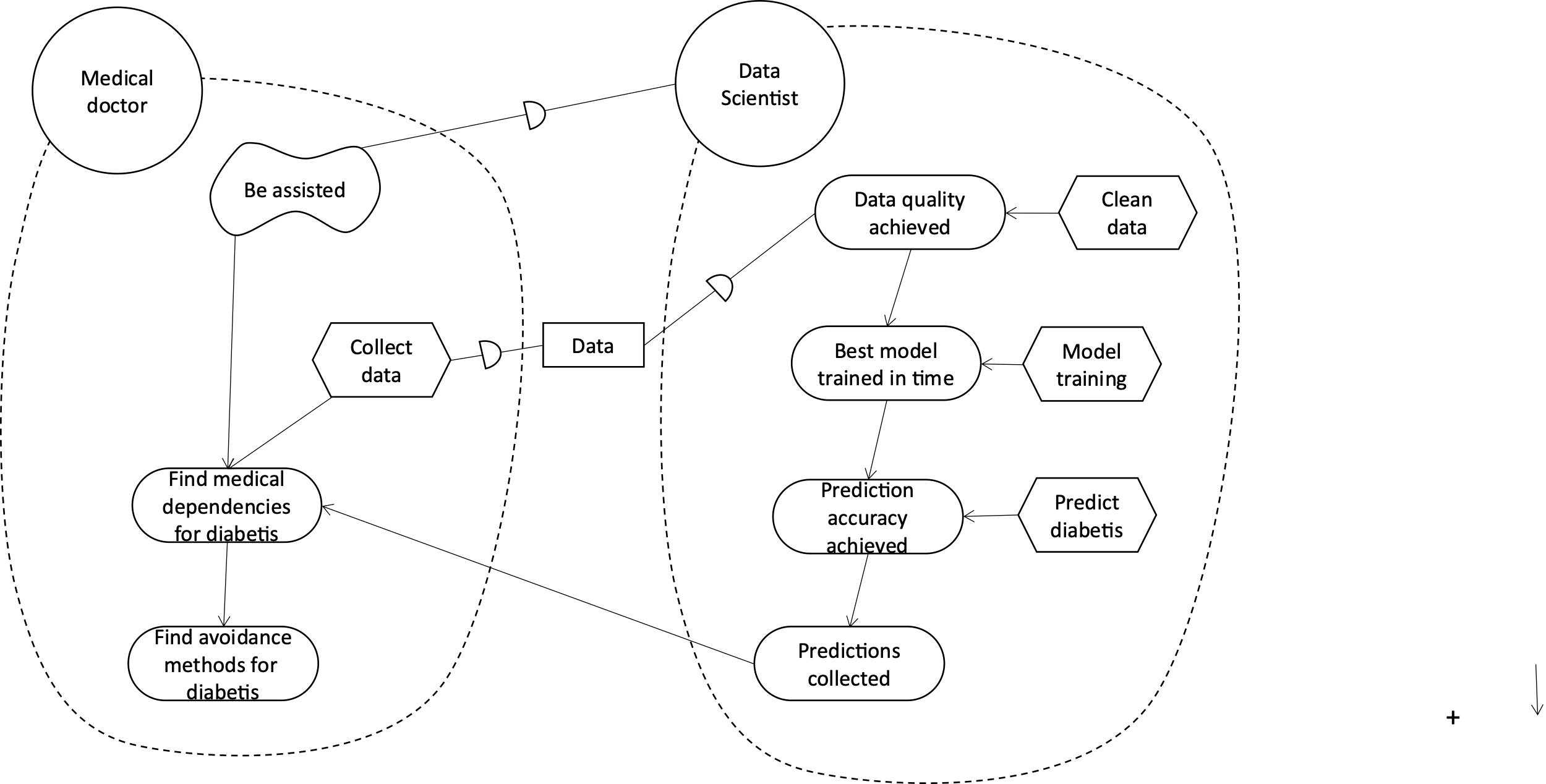}
\caption{Initial goal model.}
\label{fig:goal}
\end{figure}

For the Pima project, medical researcher and data scientist are identified as actors. An initial goal model (cf. Figure \ref{fig:goal}) states that the data scientist assists the medical researcher in achieving the goal of finding dependencies for diabetis. The data scientists’s main goal targets the collection of predictions while the medical researcher targets avoidance methods for diabetis. It mainly focusses on goals on domain level and data analytics level but abstracts from goals related to data access \cite{nyk21}. The principal-agent relationship between medical reseacher and data scientist is modeled as a soft goal ($``$Be assisted$"$). The medical doctor is responsible for collecting unbiased data and the data scientist is responsible for data quality. Several goal dependencies between actors exist in data science projects. For instance, data engineering tries to achieve data quality requirements but this also needs to comply with the bias avoidance goal of the medical doctor. Identification of goal dependencies between actors is crucial for finding ML-based solutions for domain experts.

\vspace{2\baselineskip}

\begin{figure}[H]
\includegraphics[width=15.12cm,height=7.91cm]{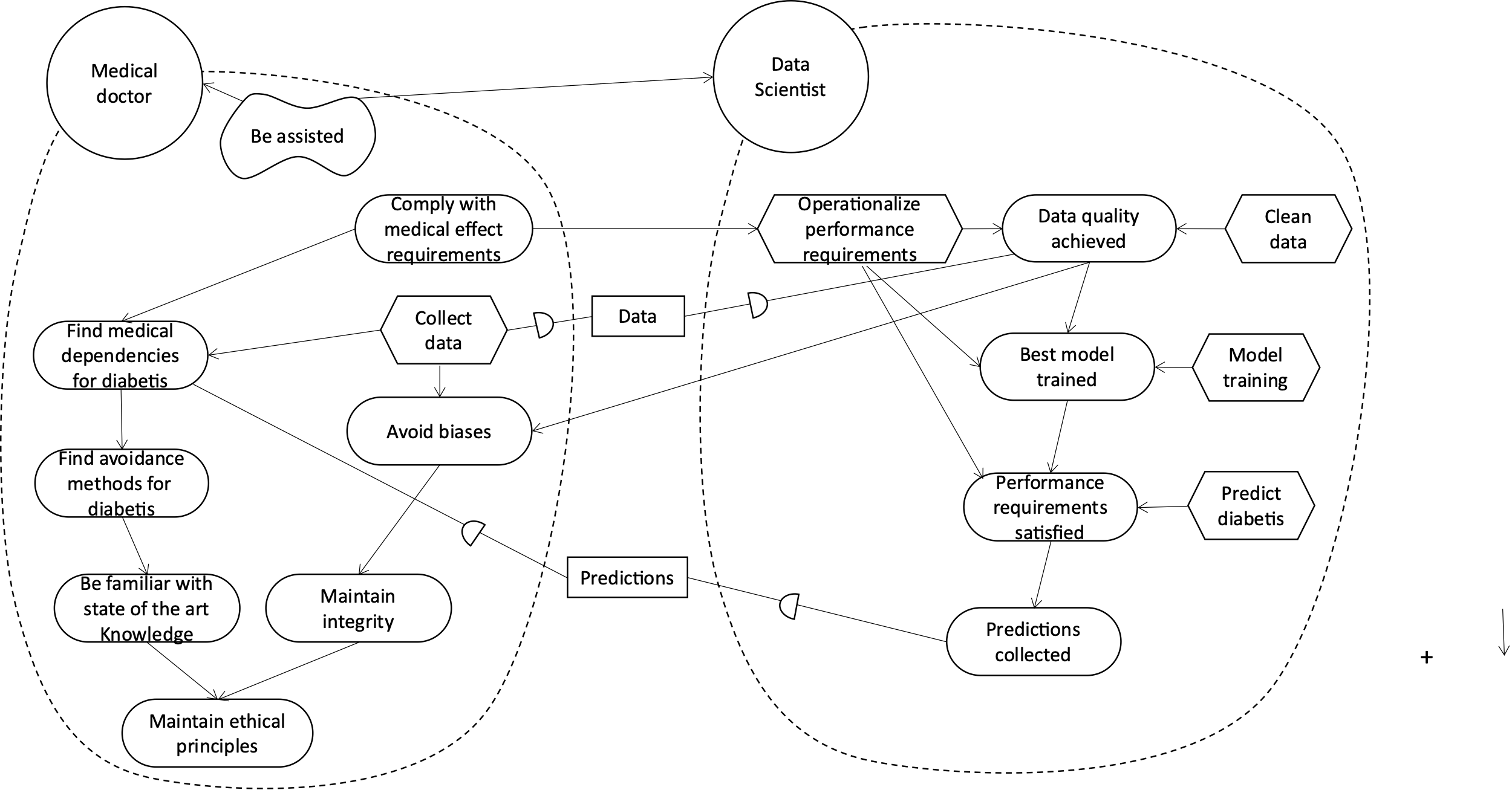}
\caption{Extended goal model.}
\label{fig:extgoal}
\end{figure}

Several issues are identified for the initial goal model after consulting literature on medical ethics \cite{h63} and discussion with medical researchers, it becomes evident that beside pure functional goals related to avoiding diabetes, medical researchers also try to follow higher ethical principles including maintenance of integrity (cf. Figure \ref{fig:extgoal}). Data scientist do generally not account for goals that drive medical researchers. Therefore, linking data quality goals of data scientists with the bias avoidance goal of medical researchers is crucial for the success of the data science project. By making this explicit, both actors become aware of this relationship and can agree on measures that support goal achievement. A similar goal relationship exists between expected effect requirements on medical side and operationalization into performance requirements. Medical researchers perceive predictions as data that becomes input. This is translated into a requirement that predictions are not provided as graphics or performance measures but as tables with input and output data. Overall, the extended goal model expresses in more detail how medical researchers and data scientist intend to collaborate that reduces misunderstanding during project implementation.

\vspace{1\baselineskip}
\textbf{Data collection} 

This dataset has been provided via Kaggle.\footnote{ https://www.kaggle.com/uciml/pima-indians-diabetes-database} The data set is accompanied with textual descriptions with units but provides no further semantical, structural, or contextual data requirements.

%

\begin{table}[]
	\begin{adjustbox}{max width=\textwidth}
		\begin{tabular}{|p{1.8cm}|p{1.8cm}|p{1.8cm}|p{1.8cm}|p{1.8cm}|p{1.8cm}|p{1.8cm}|p{1.8cm}|p{1.8cm}|} 
			\hline
			Pregnancies (number)
			&	Glucose (Plasma glucose concentration at 2 hours in an oral glucose tolerance test)	& Blood-Pressure (Diastolic blood pressure (mm Hg)
)
			& Skin-Thickness (Triceps skin fold thickness (mm)
)
			& Insulin
(2-Hour serum insulin (mu U/ml)
)
			& BMI (Body mass index (weight in kg/(height in m)2)
)
			& Diabetes-Pedigree-Function (Diabetes pedigree function
)
			& Age (years)	& Outcome (0 / 1)
\\ \hline
			6		& 	148		& 72		& 35		& 0		& 33.6		& 627		& 50		& 1
\\ \hline
			1		& 85		& 66		& 29		& 0		& 26.6		& 351		& 31		& 0
\\ \hline
			8		& 183		& 64		& 0			& 0		& 23.3		& 672		& 32		& 1
\\ \hline
			1		& 89		& 66		& 23		& 94		& 28.1		& 167		& 21		& 0
\\ \hline
			0		& 137		& 40		& 35		& 168		& 43.1		& 2.288		& 33		& 1
\\ \hline
			5		& 116		& 74		& 0			& 0		& 25.6		& 201		& 30		& 0
\\ \hline
			3		& 78		& 50		& 32		& 88		& 31		& 248		& 26		& 1
 \\ \hline
		\end{tabular}
	\end{adjustbox}
	\caption{Dataset (selection).}
	\label{tab:dataset}
\end{table}

\vspace{1\baselineskip}
\textbf{Data engineering} 

This consisted of data exploration, data preparation, and feature engineering on the dataset.

\vspace{1\baselineskip}
\textit{\uline{Data exploration}} The dataset consists of 768 cases with 7 directly collected variables, one constructed feature and one outcome variable, as shown in Table \ref{tab:dataset} and the descriptive statistics given in Table \ref{tab:descriptives}.

%

\begin{table}[]
	\begin{adjustbox}{max width=\textwidth}
		\begin{tabular}{|p{4.74cm}|p{1.24cm}|p{1.45cm}|p{1.45cm}|p{1.45cm}|p{1.25cm}|p{1.45cm}|p{1.45cm}|p{1.45cm}|} 
			\hline
		& count		& mean		& std		& min		& 25\%		& 50\%		& 75\%		& max
 \\ \hline
	Pregnancies	& 768		& 3.84		& 3.36		& 0.000		& 1.00		& 3.00		& 6.00	& 17.00
 \\ \hline
	Glucose		& 768	& 	120.89	& 	31.97	& 	0.000	& 	99.00	& 	117.00	& 	140.25	& 	199.00
 \\ \hline
	Blood¬Pressure		& 768		& 69.10		& 19.35		& 0.000		& 62.00		& 72.00		& 80.00		& 122.00
 \\ \hline
	Skin-Thickness		& 768		& 20.53		& 15.95		& 0.000		& 0.00		& 23.00		& 32.00		& 99.00
 \\ \hline
	Insulin		& 768		& 79.79		& 115.24		& 0.000		& 0.00		& 30.50		& 127.25		& 846.00
 \\ \hline
	BMI		& 768		& 31.99		& 7.88		& 0.000		& 27.30		& 32.00		& 36.60		& 67.10
 \\ \hline
	Diabetes-Pedigree-Function		& 768		& 0.47		& 0.33		& 0.078	& 	0.24		& 0.37		& 0.62		& 2.42
 \\ \hline
	Age		& 768		& 33.24		& 11.76		& 21.000		& 24.00		& 29.00		& 41.00		& 81.00
 \\ \hline
	Outcome		& 768		& 0.34		& 0.47		& 0.000		& 0.00		& 0.00		& 1.00		& 1.00
 \\ \hline
		\end{tabular}
	\end{adjustbox}
	\caption{Descriptive statistics.}
	\label{tab:descriptives}
\end{table}

\begin{center}
 \begin{figure}[H]
\centering
	\includegraphics[width=12.43cm,height=7.36cm]{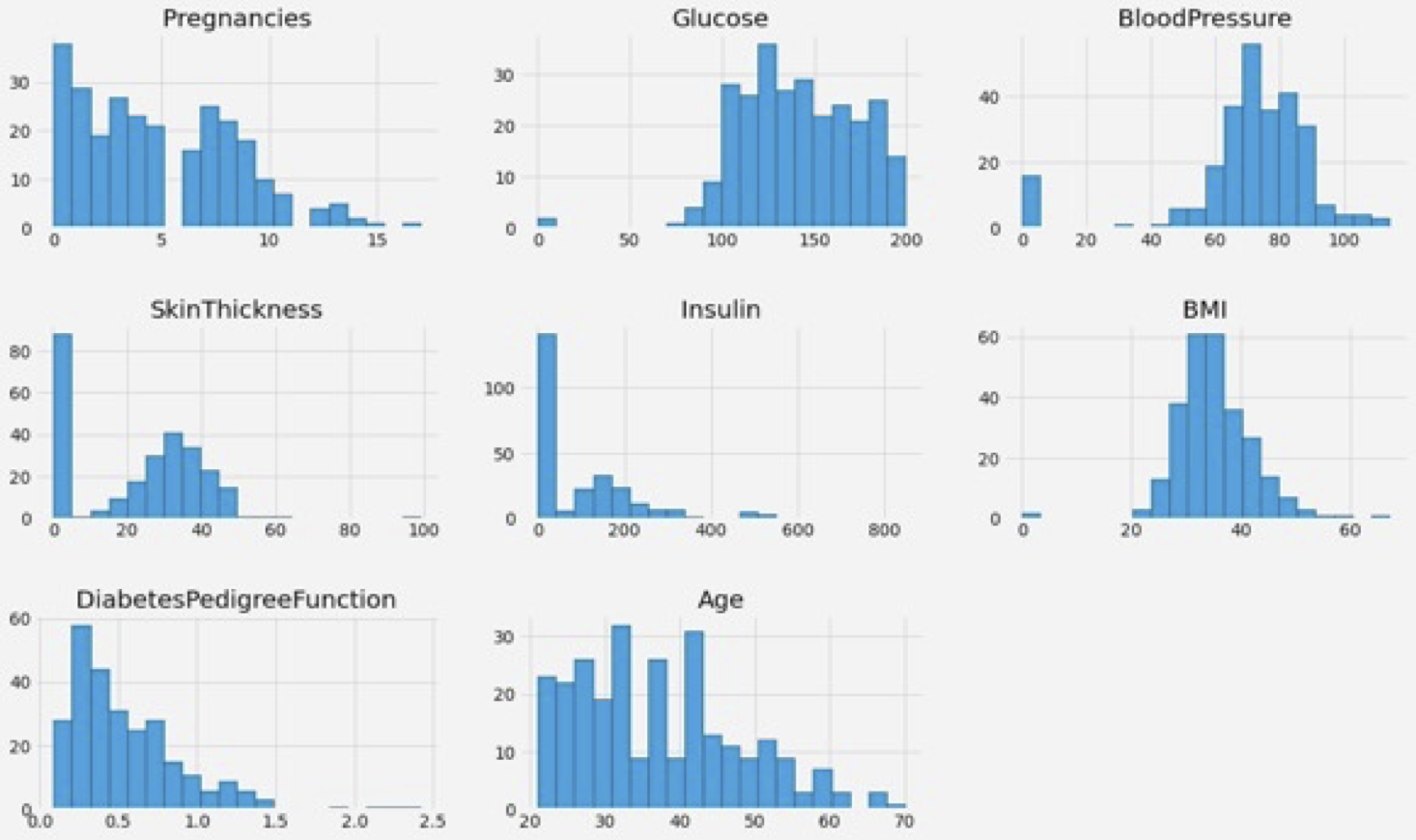}
	\caption{Probability density functions for Pima population diabetes study.}
	\label{fig:pdf}
\end{figure}

\end{center}

Visualizations of the probability density functions show that some features follow a normal distribution (blood pressure, body mass index (BMI)), others are strongly skewed (DPF, age) or indicate lower quality (insulin and skin thickness) (cf. Figure \ref{fig:pdf}).

The data requirements are summarized below. Missing data has been found for BloodPressure (4.56$\%$), SkinThickness (29.56$\%$), and Insulin (48.7$\%$). \textit{All feature values} must be positive

\begin{itemize}
	\item \textit{Num\_of\_preg (number of pregnancies) must be recorded.}

	\item \textit{Glucose (Plasma glucose concentration after 2 hours in an oral glucose tolerance test)}:  impaired glucose tolerance: between 7.8 mmol/L (140 mg/dL) and 11.1 mmol/L (200 mg/dL); levels$\geq$11.1 mmol/L at 2 hours confirm a diagnosis of diabetes.

	\item \textit{BloodPressure (diastolic blood pressure (mm Hg))}: less than 120 mm Hg

	\item \textit{SkinThickness (Triceps skin fold thickness (mm))}: no restrictions due to lack of knowledge

	\item \textit{Insulin (2-Hour serum insulin (mu U/ml))}: categorization: 1-110, 111-150, 151-240, >241 \cite{wfh84}. Interpretation and derivation of data requirements requires domain expertise\textit{.}

	\item \textit{BMI (Body mass index (weight in kg/(height in m)$\string^$2))}: ranges: <18.5 underweight, 18.5 – 24.9 normal, 25.0 – 29.9 overweight, >30 obese. Ranges do not apply for athletes.
\end{itemize}

\begin{center}
 \begin{figure}[H]
\centering
 \includegraphics[width=11.12cm,height=9.55cm]{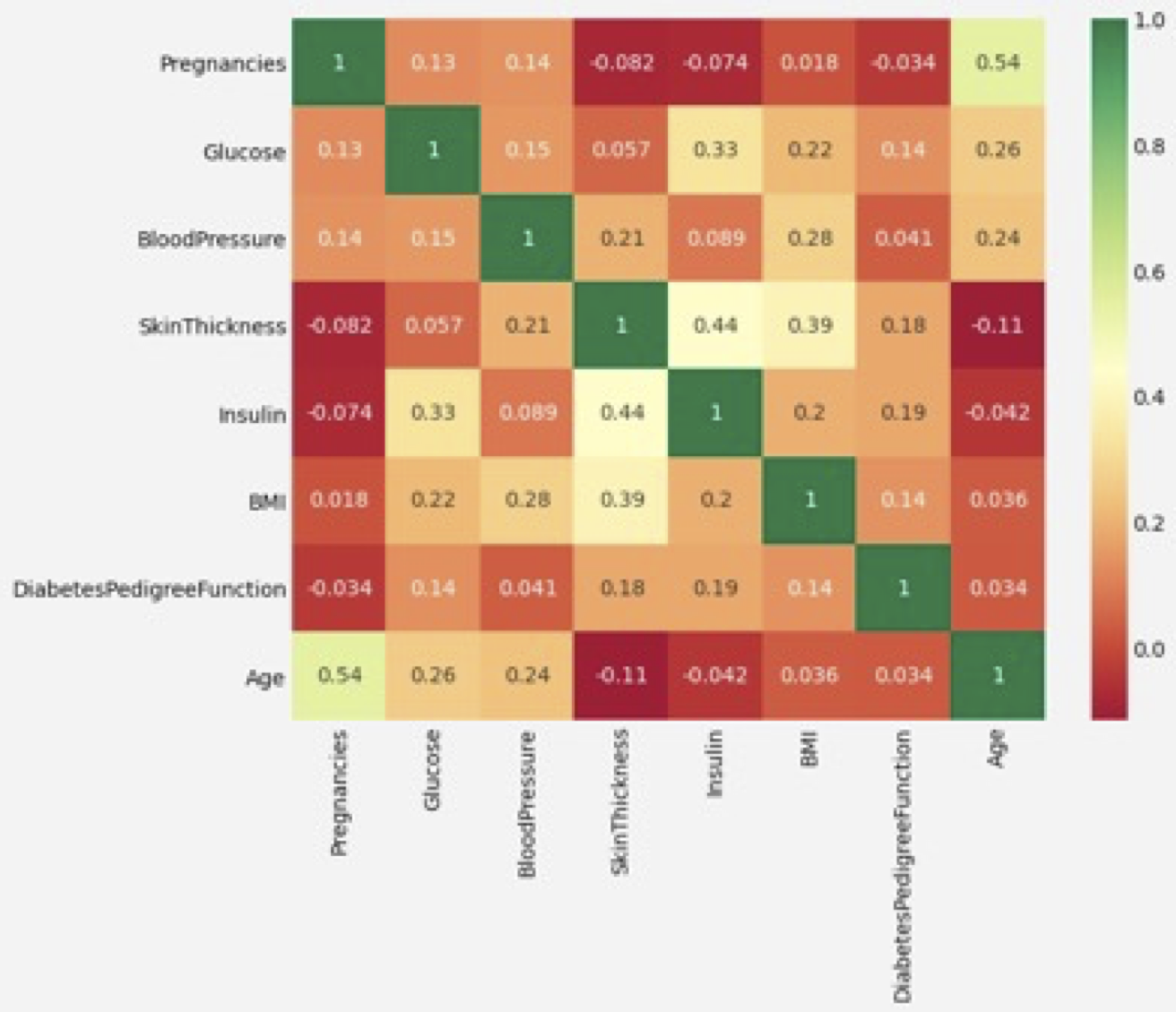}
 \caption{Correlation matrix for Pima population diabetes study.}
 \label{fig:corr}
\end{figure}

\end{center}

\begin{itemize}
	\item \textit{Age (in years)}: oldest person is less than 122 years (age of oldest person ever recorded).

\end{itemize}
The correlation matrix indicates low interactions between features. This supports the assumption that features are independent and independently contribute to estimations (cf. Figure \ref{fig:corr}).

\vspace{1\baselineskip}
\textit{\uline{Data preparation}}

At a general level, diabetes is aligned with the diabetes diagnosis ontology (DDO) \cite{ea16} that provides a rich set of concepts and relations on diabetes. DDO can be conceptually aligned \cite{es07} with the goal of the data science project; that is, the concept \textit{diabetes diagnosis} in DDO with \textit{diabetes} in the goal descriptions. Further variables can be extracted by analysis of the ontology. Ontology analysis of DDO shows that concept \textit{patient} has a high centrality degree \cite{borgatti05}, with direct connection to \textit{diabetes diagnosis }via\textit{ has diagnosis}. In order to infer independent features that can improve model performance, semantic paths can be analyzed on the basis of semantic distance in an ontology \cite{pn11}. For example, \textit{patient} is directly connected to a \textit{diabetes symptom} with 89 associated concepts. Each concept is a candidate for enhancing the dataset.

Ontology embedding involves the following mapping to SNOMED CT\footnote{http://bioportal.bioontology.org/ontologies/SNOMEDCT}.

\begin{itemize}
	\item \textit{Pregnancies: http://purl.bioontology.org/ontology/SNOMEDCT/127362006}

	\item \textit{Glucose: }http://purl.bioontology.org/ontology/SNOMEDCT/434911002

	\item \textit{BloodPressure: http://purl.bioontology.org/ontology/SNOMEDCT/75367002}

	\item \textit{SkinThickness:} \textit{http://purl.bioontology.org/ontology/SNOMEDCT/247428002}

	\item \textit{Insulin:} \textit{http://purl.bioontology.org/ontology/SNOMEDCT/67866001}

	\item \textit{BMI: http://purl.bioontology.org/ontology/SNOMEDCT/60621009}

	\item \textit{Age: http://purl.bioontology.org/ontology/SNOMEDCT/397669002}

\end{itemize}
Additionally, DDO can be used to infer additional constraints on patients. For example, patient is directly related to a demographic with 9 concepts from which invariants on social status and social relationships can be inferred for better understanding and improving the dataset. Formal ontologies are often enriched by formal axiom specifications \cite{h10}. Invariants on datasets can be derived from formal axioms by axiom mining either directly or by propagation of axioms through ontologies; that is, axioms for relational algebra (e.g., symmetry, reflexivity, and inverse), composition of relationships, sub-relationships, and part-whole relationships \cite{sm00}. This indicates that ontologies are rich sources for data exploration, improvement of data quality and data refinement. Less formal ontologies are provided by knowledge graphs that connect instance by analyzing large datasets \cite{sw13}. Because knowledge graphs are often extracted from texts by text mining \cite{gl09}, instance-connection triples:

\vspace{1\baselineskip}
 \centerline{e.g., \textit{<DFKI, locatedAt, Saarbruecken>}}
\vspace{1\baselineskip}

only provide weak support for ontological structures with concepts and relationships and, therefore, require knowledge graph refinement \cite{p17}. Because knowledge graphs resemble more social networks than ontologies, graph analytics uses techniques such as for finding centrality, communities, connectivity, and node similarity \cite{i16}, as well as rule mining \cite{h20}.

\begin{center}
 \begin{figure}[H]
\centering
\includegraphics[width=12.03cm,height=3.53cm]{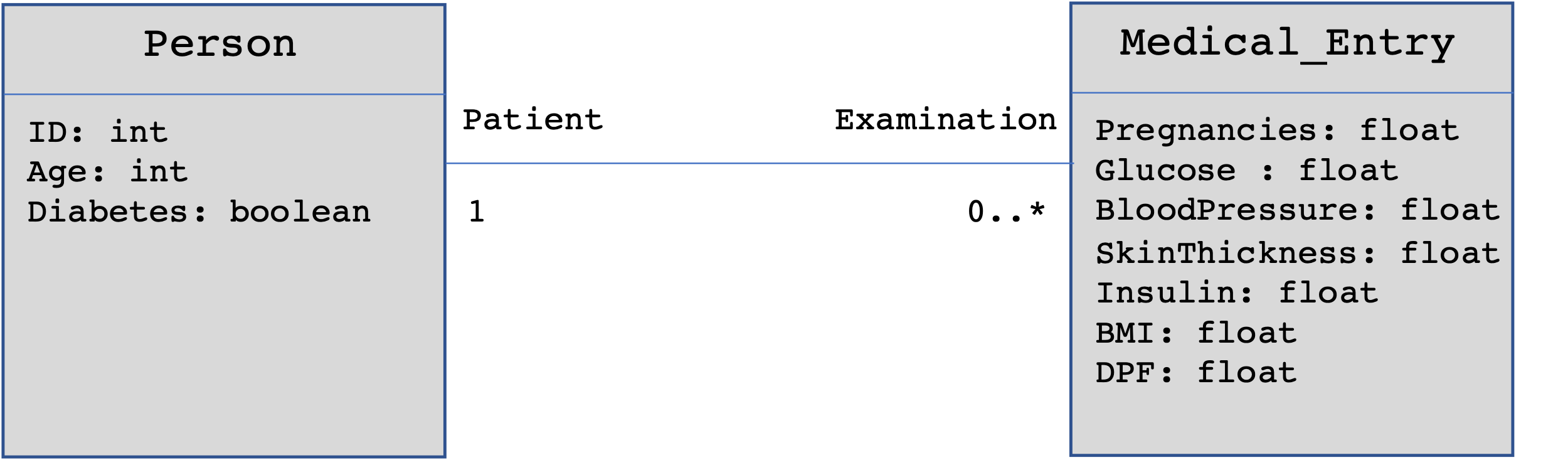}
\caption{semantic relationships between entities.}
\label{fig:semrel}
\end{figure}

\end{center}

The dataset was collected before starting this data science project. Therefore, data requirements are described ex-post and data quality is assessed instead of declaring data quality requirements. UML and OCL are potential means for describing data requirements. Data features in the data set are only connected to person via sample numbers. UML representation increased understandability by declaring a relationship between a \textit{person} and a \textit{medical entry }(Figure \ref{fig:semrel}).

Examples for object constraints (in OCL \cite{rg98}) are as follows.

\begin{itemize}
	\item \begin{flushleft}
C1: $``$Feature values of a person cannot be zero or negative$``$ \\ context Person \\ inv: self.Examination->forAll(e $\vert$ e.Glucose > 0)
\end{flushleft}

	\item \begin{flushleft}
C2: $``$A patient‘s age is between 15 and 120$``$ \\ context Person \\ inv: self.Patient.age >$=$ 15 and self.patient.age < 120
\end{flushleft}

	\item \begin{flushleft}
C3: $``$A patient has diabetes if Glucose is above 200$``$ \\ context Person \\ inv: self.Patient.Glucose >$=$ 200 -> self.Patient.Diabetes$=$1
\end{flushleft}

\end{itemize}

Given these data requirements, this dataset violates constraint C1 because it contains values of 0 for all features, but satisfies constraint C2. Constraint C3 raises an issue because no glucose value is above 200mg/dl, a strong indicator for diabetes.

Several data quality exist within this dataset. Glucose data values range from 0 to 199. This finding contradicts the data requirement that diabetes is diagnosed with a diabetes >200mg/dl. Analyzing the data collection procedure \cite{wfh84} shows that all cases have been deleted with diabetes occurring within one year of an examination. However, the exact cut at 199mg/dl suggests that, instead, all values $\geq$200mg/dl were deleted regardless of subsequent progression (cf. Figure \ref{fig:pdf}). Furthermore, skin thickness and insulin are unreliable predictors due to lack of data. This is shown in Table \ref{tab:dq}.

%

\begin{table}[ht]
	\begin{adjustbox}{max width=\textwidth}
		\begin{tabular}{|p{2.71cm}|p{2.08cm}|p{2.23cm}|p{1.69cm}|p{1.73cm}|p{1.96cm}|p{1.52cm}|p{1.22cm}|p{1.14cm}|} 
			\hline
			& & Number of pregnancies & Glucose	 & Blood Pressure	 & Skin Thickness	 & Insulin	 & BMI	 & Age \\ \hline
			\multirow{7}{2.71cm}{Accuracy} &	Believability &		++	 &	0	 &	++	 &	++	 &	++	 &	++	 &	++
\\ \cline{2-9}
			& Accuracy	 &	++	 &	--	 &	++	 &	++	 &	++	 &	++	 &	++
\\ \cline{2-9}
			 &	Objectivity	 &	++	 &	0	 &	++	 &	++	 &	++	 &	++	 &	++
\\ \cline{2-9}
			 &	Completeness	 &	++	 &	--	 &	++	 &	-	 &	-- &	++	 &	++
\\ \cline{2-9}
			 &	Traceability	 &	0	 &	0	 &	0	 &	0	 &	0	 &	0	 &	0
\\ \cline{2-9}
			 &	Reputation	 &	++	 &	++	 &	++	 &	++	 &	++	 &	++	 &	++
\\ \cline{2-9}
			 &	Variety	 &	0	 &	0	 &	0	 &	0	 &	0	 &	0	 &	0
\\ \hline
			\multirow{6}{2.71cm}{Relevancy}	& Value-added		& ++		& 0		& ++		& -		& -		& ++		& ++
\\ \cline{2-9}
				& Relevancy		& ++		& ++		& ++		& +	+		& ++		& ++
\\ \cline{2-9}
				& Timeliness		& 0		& 0		& 0		& 0		& 0		& 0		& 0
\\ \cline{2-9}
				& Ease of operation		& ++		& ++		& ++		& 0		& 0		& ++	& 	++
\\ \cline{2-9}
				& Appropriate amount of data		& +		& +		& +		& --		& --		& +		& +
\\ \cline{2-9}
				& Flexibility		& +		& +		& +		& +		& +		& +		& +
\\ \hline
				\multirow{4}{2.71cm}{Representation}	& Interpretability		& ++		& ++		& ++		& ++		& ++		& ++		& ++
\\ \cline{2-9}
				& Ease of understanding		& +		& +		& +		& +		& +		& +		& +
\\ \cline{2-9}
				& Consistency		& +		& +		& +		& 0		& 0		& +		& +
\\ \cline{2-9}
				& Conciseness		& +		& +		& +		& +		& +		& +		& +
\\ \hline
			\multirow{3}{2.71cm}{Accessibility} &	Accessibility	 &	++	 &	++	 &	++	 &	++	 &	++	 &	++	 &	++
\\ \cline{2-9}
			 &	Cost-effectiveness	 &	++	 &	++	 &	++	 &	++	 &	++	 &	++	 &	++
\\ \cline{2-9}
			 &	Access security	 &	++	 &	++	 &	++	 &	++	 &	++	 &	++	 &	++
\\ \hline
		\end{tabular}
	\end{adjustbox}
	\caption{Data quality assessment of dataset.}
	\label{tab:dq}
\end{table}

Missing data is important for this dataset. For healthcare data, a popular imputation method is $``$multiple imputation using chained equations$"$ (MICE) \cite{w13}. Simpler strategies are replacement by mean, median, or most frequent values. It is interesting to note that the most popular solution for this dataset on Kaggle (45,000 views out of 1067 unique solutions) uses a mix of mean and median without providing justification for doing so.

\begin{itemize}
	\item Value 0 for \textit{Glucose}, \textit{BloodPressure}, \textit{SkinThickness}, \textit{Insulin}, \textit{BMI} are replaced by \textbf{NAN}

	\item \textit{Glucose, BloodPressure} NAN values replaced by \textbf{mean}

	\item \textit{SkinThickness, BMI} NAN values replaced by \textbf{median}

\end{itemize}
\textit{\uline{Feature engineering}}

With the \textit{Diabetes Pedigree Function} (DPF), this dataset also provides an engineered feature. In machine learning development projects this is either: provided by domain experts; or created during feature engineering in collaboration between domain experts and data scientists and then added to the dataset. Domain experts defined a Diabetes Pedigree Function (DPF) that provides a synthesis of the diabetes mellitus history in relatives and the genetic relationship of those relatives to the subject. The DPF uses information from parents, grandparents, full and half siblings, full and half aunts and uncles, and first cousins. It provides a measure of the expected genetic influence of affected and unaffected relatives on the subject's eventual diabetes risk \cite{wfh84}:

\begin{equation}
DPF =  \frac{\sum_{i}^{}K_{i}\left(88 - ADM_{i}\right) + 20}{\sum_{j}^{}K_{j}\left(ADM_{i} - 14\right) + 50}
\end{equation}

i: all relatives i who had developed diabetes by the subject’s examination date

j: all relatives j who had not developed diabetes by the subject’s examination date

K\textsubscript{x}: percent of genes shared by relative x and set at:

\begin{itemize}
	\item 0.5 when the relative x is a parent or full sibling

	\item 0.25 when the relative x is a half sibling, grandparent, aunt or uncle

	\item 0.125 when the relative x is a half aunt, half uncle \textbf{or} cousin

\end{itemize}
ADM\textsubscript{i}: age in years of relative i when diabetes was diagnosed

ACL\textsubscript{j}: age in years of relative at the last non-diabetic examination

88 / 14: maximum and minimum ages at which relatives developed DM

20 / 50: moderating constants

This definition provides semantical and structural data requirements for the DPF variable thoroughly embedded into the domain of diabetes research (contextual data requirements). This definition could be developed into formal representations that support re-use and merging this dataset in other ML model training. Similar domain knowledge exists for glucose, blood pressure, skin thickness, and insulin.

\vspace{1\baselineskip}
\textbf{Model training. } 

Model training is governed by performance requirements as constraints for functional requirements. For building trust in medical treatment, prediction accuracy should be high with a high sensitivity (recall) value close to 100$\%$; i.e., the percentage of missed positives should be small. Slightly less important is specificity (true negative rate); that is, not many people should receive treatment, even though they are healthy. These tradeoffs and thresholds must be based on domain knowledge. Reported sensitivity (0.78) and specificity scores (0.77) are taken as reference values.

At the beginning, a series of machine learning model types is trained and evaluated by applying default hyperparameter values. Accordingly, Gradient boosting performs best with an accuracy of more than 88$\%$, recall (sensitivity) 87$\%$, precision 88$\%$ and F1 value 88$\%$ \cite{lt03}. Model training depends on selecting the best ML model type and leveraging domain knowledge. Selecting the best ML model type is the core of any model training phase, and requires application of different models and validation of model performance. Leveraging domain knowledge depends on the interaction between domain experts and data scientists. Domain knowledge leveraging supports: knowledge to data or data to knowledge. The first direction, \textit{knowledge to data}, is used if a domain expert can express domain knowledge in a way that can be transformed into additional features. For instance, if a person is younger than 30 years with a plasma glucose level <120 mg/dl, then she is less likely to suffer from diabetes in the next 5 years. This heuristic rule can be expressed by an OCL rule with an added binary feature $kl1$ initialized with $0$ for all samples:

\vspace{1\baselineskip}
\texttt{context Person}\\
\texttt{inv: self.Patient.age < 30 and}\\
\tab \quad\texttt{self.medical\_entry.glucose < 120}\\
\tab \quad\texttt{-> forAll(e $\vert$ e.kl1 = 1)}

\vspace{1\baselineskip}
For \textit{data to knowledge,} a data scientist analyzes the dataset and attempts to extract heuristic rules that will subsequently be evaluated by domain experts. For instance, if a data scientist finds support for a hypothetical rule that younger women with fewer pregnancies are less likely to suffer from diabetes in the coming years, then this is expressed as an OCL rule:

\vspace{1\baselineskip}
\texttt{context Person}\\
\texttt{inv: self.Patient.age < 30 and}\\
\tab \quad\texttt{self.medical\_entry.pregnancies <$=$ 6}\\
\tab \quad\texttt{->forAll(e $\vert$ e.kl2 $=$ 1)}

If rule $kl2$ is verified by domain experts, another binary variable is added to the dataset. From the \textit{data to knowledge} strategy, an additional 16 binary features were found and added to the dataset.\footnote{https://www.kaggle.com/vincentlugat/pima-indians-diabetes-eda-prediction-0-906} These heuristic rules increased model performance from an accuracy of 0.73 for gradient boosting to 0.89 with a recall of 0.84 and a precision of 0.86. That is an increase of more than 20$\%$ above the initial model performance.

The following ML model types were trained as shown in Table \ref{tab:modeltypes}.

%

\begin{table}[]
	\begin{adjustbox}{max width=\textwidth}
		\begin{tabular}{|p{4.18cm}|p{1.95cm}|p{1.22cm}|p{3.44cm}|p{1.91cm}|p{1.06cm}|} 
			\hline
			\textbf{Model} &	\textbf{Accuracy} &		\textbf{AUC}	 &	\textbf{Recall / Sensitivity}	 &	\textbf{Precision}	 &	\textbf{F1}
\\ \hline
			Light Gradient Boosting	 &	0.89	 &	0.94	 &	0.84	 &	0.86	 &	0.85
\\ \hline
			Gradient boosting	 &	0.89	 &	0.95	 &	0.81	 &	0.85	 &	0.83
\\ \hline
			Logistic regression	 &	0.84	 &	0.91	 &	0.73	 &	0.78	 &	0.76
\\ \hline
			Support vector classifier &		0.85	 &	0.91	 &	0.75	 &	0.81	 &	0.78
\\ \hline
			Decision tree &		0.86	 &	0.81	 &	0.82	 &	0.79	 &	0.81
\\ \hline
			K nearest neighbors &		0.80	 &	0.88	 &	0.59	 &	0.77	 &	0.67
\\ \hline
		\end{tabular}
	\end{adjustbox}
	\caption{Model types.}
	\label{tab:modeltypes}
\end{table}

\vspace{1\baselineskip}
\textbf{Model optimization. } 

Model optimization is also governed by performance requirements. Most machine learning model types have hyperparameters, such as the number of neighbors$k$ that are used in KNN. Finding an optimal set of hyperparameters is a NP-complete search problem.\footnote{Thank you to an anonymous reviewer for this suggestion.} Even if an optimal set of hyperparameters could be computed, it is not possible to assess whether a viable solution has been identified because the solution might suffer from inductive fallacy.  Best practices can be expressed by heuristic rules or knowledge graphs. Model optimization is a technical task similar to optimization of a database system by configuration of database management parameters. More research is needed to understand the impact on domain knowledge and on data science knowledge for optimizing ML models.

\vspace{1\baselineskip}
\textbf{Model integration. }

After finalizing the ML model, it is integrated into the information system. Validation procedures can be used to assess compliance with business requirements, goal requirements, data requirements, legal requirements, ethical requirements, and functional and non-functional requirements. Field tests on newly collected data are used to build trust in information system performance. Empirical studies on information systems adoption, usability, cost effectiveness and other non-functional requirements are used for practical evidence of ML development results. The diabetes machine learning model was not integrated into an information system. Therefore, model integration is not relevant in this example.

\vspace{1\baselineskip}
\textbf{Analytical decision making. }

The value of a diabetes information system lies in its potential to support medical workers. By sampling new data, medical workers receive predictions on the risk of patients suffering from diabetes in the future so that countermeasures can be recommended, even in real-time.

\section{Machine Learning for Conceptual Modeling}

Machine learning can also contribute to conceptual modeling. Many of the challenges of applying machine learning to conceptual modeling deal with knowledge generation. Here, the term \textit{knowledge} refers generally to the constructs of a conceptual model. Knowledge challenges can be organized into three categories: incomplete knowledge, incorrect knowledge, and inconsistent knowledge. Incomplete knowledge includes missing or limited entities and/or relationships. Incorrect knowledge includes incorrect entity and relationship labels or incorrect facts (e.g., cardinalities). Inconsistent knowledge includes different labels for the same entity or merging entities with the same labels.

The first, and probably easiest to understand, is missing entities or relationships. Knowledge could be extracted to identify where there is incompleteness in modeling of an application domain, or potential missing relationships, which would require interaction between a domain expert and a conceptual modeling expert.

It is possible to infer what concepts and synonyms are extracted from a text. Then, the potential entity concepts can be used to create a graph that might indicate missing pieces or something that is incorrectly labeled (wrong entity label recognition). There could also be inconsistent relationships, or the potential to make incorrect inferences. Such basic kinds of research challenges are well-known. However, anchoring such inferences in knowledge graphs could support the combining of research on knowledge graphs with data analytics and conceptual modeling. We can consider analytics on text and how to extract conceptual modeling-like structures from it, as well as rule and graph mining.

\begin{itemize}
	\item \textit{General supervised learning}: linear regression, support vector machines, decision trees, random forests, boosting models, multi-layer perceptrons, deep neural networks \cite{schmidhuber15}

	\item \textit{Sequence learning}: recurrent neural networks (incl. LSTM) \cite{hs97}

	\item \textit{Generative learning}: generative adversarial neural networks \cite{g14}

	\item \textit{Graph learning}: graph neural networks, recurrent graph neural networks, convolutional graph neural networks (\cite{w20})

	\item \textit{Unsupervised learning}: KNN, k-means, PCA \cite{htf09}

	\item \textit{Reinforcement learning}: dynamic learning of agents through rewards gained from actions in environments \cite{sb18}.

\end{itemize}
Extracting knowledge structures from datasets by means of machine learning is a fast growing research field. Table \ref{tab:extraction} provides an non-exhaustive overview of using machine learning models for extracting different conceptual model structures. Associative rules is a robust technology for extracting relational knowledge. Approaches for estimating relationships between entities as link predictions are more sophisticated. Process discovery based on analysing log files is another promising area of research. These approaches, however, do not consider semantics. This might be why ontology extraction using machine learning is still restricted to ontology matching and mapping, although there are successful language translation systems that do not have explicit semantic representations \cite{wu16}.

%

\begin{table}[]
	\begin{adjustbox}{max width=\textwidth}
		\begin{tabular}{|p{2.99cm}|p{3.0cm}|p{3.5cm}|p{3.5cm}|p{3.5cm}|} 
			\hline
			\textbf{ML $\rightarrow$ CM}	& \textbf{Rules, anomalies and explanations}	& 	\textbf{Semantic models }
			(ERM, i*)		& \textbf{Ontologies}		& \textbf{Process models}
(EPC, BPMN)
 \\ \hline
			General Supervised learning		& Associative rules \cite{ais93}; rule extraction \cite{bb10}
	& 	& Ontology mapping and matching \cite{d04, nso11}
	& Process discovery \cite{augusto18}; event abstraction \cite{v20}

 \\ \hline
			Sequence learning	& 	Rule extraction \cite{ms17}
	& Named entity recognition \cite{cn16}; link prediction \cite{c19}
	& Ontology matching \cite{jx20}
	&  \\ \hline
			Generative learning		& 	& 	Link prediction (Qin et al. 2020)	& 	&
 \\ \hline
			Graph learning		& 	& 	Link prediction \cite{h06, dettmers18} Relational learning \cite{ntk11} 	& 	&  \\ \hline
			Unsupervised learning	& 	Anomaly detection  \cite{a17}
	& Link prediction \cite{llc10}
	& Concept learning Mi et al. 2020) 	& 	Event abstraction \cite{v20}

 \\ \hline
			Reinforcement learning	& 	Rule extraction \cite{pv20}
	& 	& 	& \\ \hline
		\end{tabular}
	\end{adjustbox}
	\caption{Extraction of conceptual models with machine learning.}
	\label{tab:extraction}
\end{table}

\vspace{13\baselineskip}
\subsection{Text mining}

Text mining, which discoveres conceptual structures from unstructured sources \cite{w04},  became popular with the increasing use of social media, such as Facebook and Twitter. Various natural language processing (NLP) methods are available for filtering keywords based on domain knowledge and domain lexica. Preprocessing of textual data removes stopwords and reduces words to word stems. Shallow parsing identifies phrases and recognizes named entities with ontology mappings \cite{bsw99}. At a semantic level, word sense disambiguation and identification of negations are processed. Negation is difficult to deal with because it might mean that entities, relationships, and larger conceptual structures are excluded, or do not exist. Entity linking by NLP methods leverage ontologies \cite{whh18}. Particularly challenging is resolving references given by anaphora or by spatio-temporal prepositions. Text mining is a specific type of data mining that focuses on unstructured text. Mining association rules \cite{ais93} is often used for extracting heuristic rules. Automatic entity classification by ML models, in particular, decision trees, is another rule mining technique \cite{hk00}.

\subsection{Knowledge graphs}

Machine learning has focused on extracting latent representations from euclidean data including images, text and videos. The need to process graph data has also important. Graph structures are natural representations in domains, such as e-commerce \cite{y19, x20}, drug discovery \cite{c18} and chemistry \cite{coley19}, production and manufacturing, supply chain management \cite{l17} and network optimization \cite{r20}. Graphs are a natural means for explanation of opaque machine learning models. Therefore, they are used as input to machine learning and output from machine learning, providing an important perspective for how machine learning can support conceptual modeling. A knowledge graph $``$(i) describes real world entities and their interrelations, organized in a graph, (ii) defines possible classes and relationships of entities in a schema, (iii) allows for potentially interrelating arbitrary entities with each other; and (iv) covers various topical domains.$"$ \cite{p17}.

Adding typed links between data from different sources is an active research topic on semantic technologies \cite{bhb11}. Google, for example, extended semi-automatic annotation procedures by automatic extraction of knowledge graphs \cite{singhal12} based on existing sources, such as dbpedia \cite{a07}, YAGO \cite{skw07} or WordNet \cite{m95}.

\begin{figure}[H]
\includegraphics[width=11.99cm,height=6.44cm]{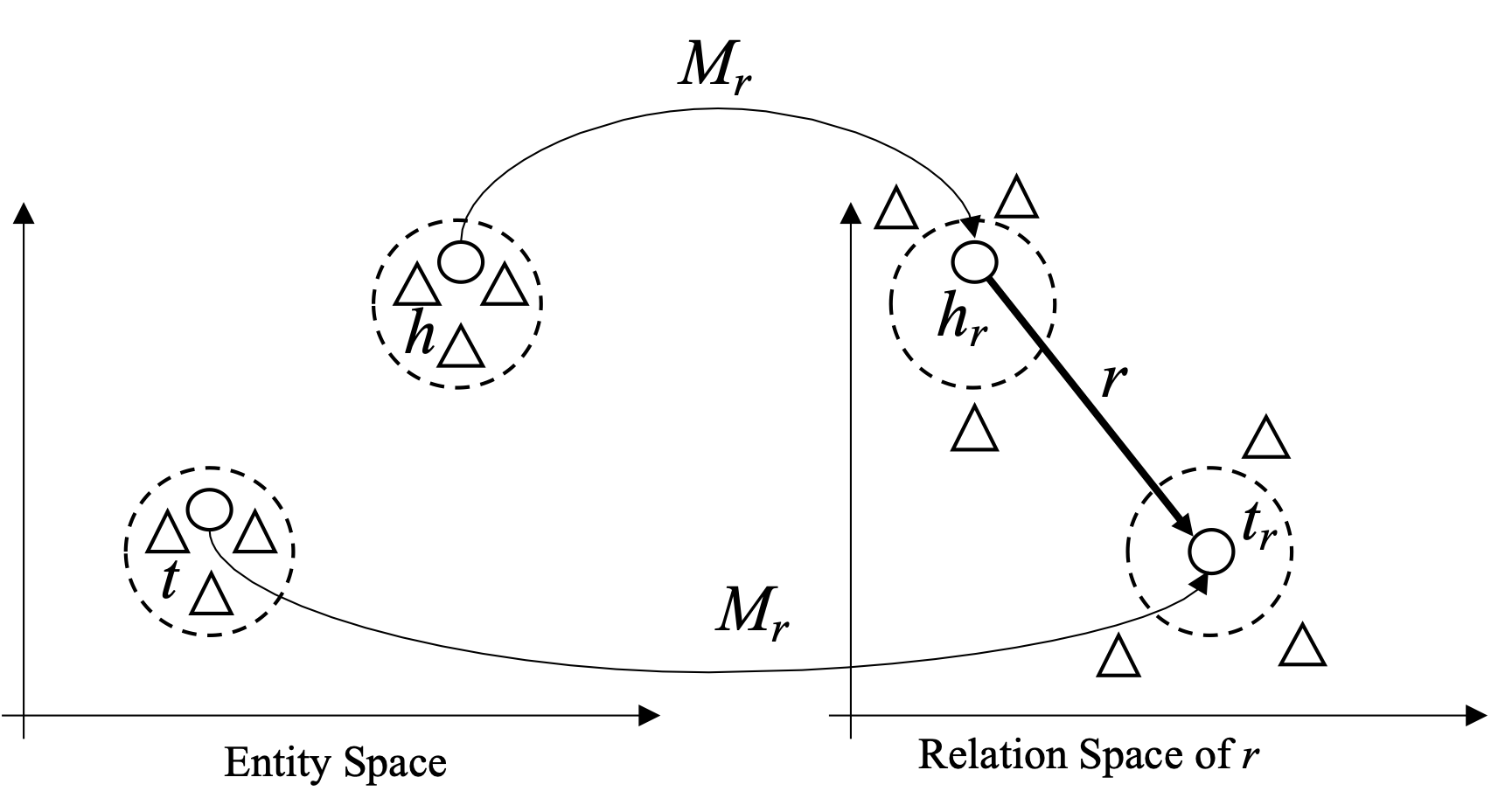}
\caption{Link prediction with TransR. Adapted from \cite{l15}.}
\label{fig:linkpred}
\end{figure}


Knowledge graphs extract named connections between instances, called \textit{link prediction between entities} \cite{l15}, which provide initial support for partial conceptual models. With large datasets, the quality of triples found by knowledge graphs is often low; that is, many links are tautological, or even meaningless. Entity resolution, collective classification, and link prediction can be used to construct consistent knowledge graphs based on probabilistic soft logic \cite{p13}. A standard approach is to transform data into vector spaces resembling principal component analysis (PCA). For instance, TransR proposed by Lin et al. \cite{l15} projects entities $h$ (head) and$t$ (tail) from an entity space into a r-relation space by a mapping function $M_{r}$ that supports $h_{r} + r \approx t_{r}$ and, thus, finds a set of entities $t_{r}$ that fulfil a triple $<h, r,t>$ (cf. Figure \ref{fig:linkpred}). $M_{r}$ is a vector embedding model that is trained on a loss function that minimizes a distance $d(h + r,t)$ of ground truth and $d\left(h + r,t\right)$ estimations. ML models based on embeddings are generative models (cf. section 3.1.2) that encode entities and relationships in vector spaces, make predictions into the input space, called \textit{decoding}, and measure the reconstruction error as an indicator for model performance. Hence, graph embeddings are powerful models used for various graph analytical tasks.

\textit{Graph analytics} use social network theory by analyzing distances and directed connections (ties) of knowledge graphs to support semantic annotations, such as similarity, centrality, community, and paths \cite{wf94}. Similar to data engineering, knowledge graphs are explored and missing elements predicted (e.g., link prediction, completion and correction) \cite{k07}. When using knowledge graphs to support machine learning, graphs are embedded into multidimensional vector spaces by preserving a proximity measure defined on knowledge graph,$G$ \cite{gf18}. Graph embedding (GE) reduces the dimensionality of a graph. Currently, autoencoder models are used for embedding graph nodes and preserving non-linear dependencies \cite{wcz16}. Graph embeddings are used for link prediction and node classification and, thus, can be indirectly used for concept classification, identification of relationships, and ontology learning \cite{mlr18}. There is a growing interest in graph neural networks, which consider graphs as input, instead of tabular data. Graph neural networks (GNN) are extensions of graph embedding with emphasis on deep learning architectures, such as recurrent neural networks, convolutional neural networks and autoencoders \cite{w16}.\ \

Graph embeddings (GE) and graph neural networks (GNN) require graph input that meets  quality requirements. Applications based on GE and GNN need to satisfy all requirement types, in the ML development cycle. This makes conceptual modeling methods and tools important assets for the development of GE/GNN-based information systems. Research on knowledge graphs provide important technologies for research on linked data and ontologies in general \cite{h20} and conceptual modeling in particular, including reasoning and querying over contextual data, and rule and axiom mining.

\section{Summary and roadmap}

In this paper, we  align conceptual modeling and machine learning in both directions. Due to the early stages of research related to this pairing, challenges remain. Conceptual modeling was motivated by research on relational databases \cite{c76} and procedural programming whereas machine learning is a child of statistics and linear algebra. Although it is clear that conceptual modeling can support data management of machine learning, many challenges remain for  supporting the development of model architectures, model training, model testing, model optimization, deployment, and maintenance in information systems. For example, deep convolutional neural network architectures consist of multiple layers with different sizes taking different roles, such as convolution and pooling layers, and application of different convolutional kernels \cite{ksh12}.

Many technical research issues emerge including:

\begin{itemize}
	\item Use of data ontologies and design patterns for data engineering

	\item Alignment of data engineering with databases and big data stores

	\item Models for mining data streams

	\item Design patterns for model architectures

	\item Process models for model development

	\item Performance models for model development.

\end{itemize}
Because ML models are central to services delivered by information systems, they also require alignment with enterprise architectures. Research issues include the following:

\begin{itemize}
	\item Frameworks for aligning model architectures and service architectures and enterprise architectures

	\item Frameworks for alignment of performance metrics and key performance indicators.

\end{itemize}
Decision making, which relies on machine learning must consider:

\begin{itemize}
	\item The quality and performance models for data-driven decision making (cf. \cite{pf13})

	\item Conceptual modeling in real-time data-driven decision making with batch and streaming data.

\end{itemize}
Conceptual modeling is well positioned when it comes to structuring requirements for complex systems, including ML-based systems. Recent proposals structure data science development by views and goal models \cite{l19, nyk21}. In the direction of machine learning for conceptual modeling, research issues are in their infancies. Knowledge graphs extracted from data are promising areas with clear connections to conceptual modeling. Merging knowledge graphs with formal ontologies is a challenging, but also promising, research topic.

\section{Conclusion}

The fields of machine learning and conceptual modeling have been active areas of research for a long time, making it reasonable to expect that it might be advantageous to explore how one might complement the other. This paper has identified possible synergies between the machine learning and conceptual modeling and proposed a framework for conceptual modeling for machine learning. Conceptual{\scriptsize  }modeling can be helpful in supporting the design and development phases of a machine learning-based information system. Feature sets, especially, must be consistent so that data scientists can create valid solutions. This can be best accomplished by including domain knowledge, as represented by conceptual models. Inversely, machine learning techniques are very successful at obtaining or scraping large amounts of data, which can be used to identify concepts and patterns that could be useful for inclusion in conceptual models. There are, of course, many challenges related to incorrect, incomplete, or inconsistent knowledge. Nevertheless, it is feasible to pair conceptual modeling with machine learning. This paper has identified some of the challenges inherent in achieving this pairing in an attempt to lay the groundwork for future research on combining conceptual modeling and machine learning.

\vspace{2\baselineskip}
{\large \textbf{Acknowledgements}}

{\large This paper was based on a keynote presentation given by the first author at the \textit{International Conference on Conceptual Modeling}. The authors wish to thank Peter Chen and Carson Woo, and Oscar Pastor for their support of this paper, Iaroslav Shcherbatyi for sharing his technical expertise in machine learning and Michael Schrefl for identifying this topic. We also thank the anonymous reviewers and Roman Lukyanenko for provided valuable insights and comments.   \par}


%

\providecommand{\bysame}{\leavevmode\hbox to3em{\hrulefill}\thinspace}
\providecommand{\MR}{\relax\ifhmode\unskip\space\fi MR }
\providecommand{\MRhref}[2]{%
	\href{http://www.ams.org/mathscinet-getitem?mr=#1}{#2}
}
\providecommand{\href}[2]{#2}

\end{document}